# Modelling the transition from a socialist to capitalist economic system

Ivan O. Kitov

## Abstract

The transition of several East and Central European countries and the countries of the Former Soviet Union from the socialist economic system to the capitalist one is studied. A recently developed microeconomic model for the personal income distribution and its evolution and a simple functional relationship between the rate of the per capita GDP growth and the attained level of the per capita GDP are used to describe the transition process. The developed transition model contains only three defining parameters and describes the process of real GDP per capita evolution during the last 15 years. It is found that the transition process finished in the Central European countries several years ago and their economic evolution is defined by pure capitalist rules. In the long run, this means that the future of these countries has to follow the same path, i.e. dependence on the per capita GDP growth rate of the per capita GDP itself, as the developed countries have had in the past. If the best GDP evolution scenario occurs for the studied countries, they will be able to maintain the absolute lag in per capita GDP relative to most developed countries including the USA. But they will never catch the advanced countries if they follow the same rules of development. In Russia and some countries of the Former Soviet Union the transition process is still far from complete.

## Introduction

A microeconomic model for the personal income distribution and its evolution in time is developed in [1]. When aggregated over the population above 15 years of age, the model transforms into a macroeconomic model describing evolution of GDP and per capita GDP in the USA. The model characterizes the capitalist system which has no formal limit to personal income. The limitation on personal income is a characteristic feature of the socialist system and explains its relatively lower GDP growth rate compared to that of the capitalist system [2].

The socialist system has undergone disintegration during the last 15 years (since 1989). This period is often called the transition from socialism to capitalism. There are some specific features that are different from those observed in a pure socialist or pure capitalist system characterizing the period. At any moment between the start of the process and its end the economic state is, supposedly, not just a mechanical sum of the socialist and capitalist fractions. Analogously, such a simple mixture of two states is observed in phase transition processes (like ice/water phase transition at 0C$^o$) as determined by physical laws of mass and energy conservation. There must be, however, some interaction between the processes of the socialist system disintegration and the capitalism build-up. Some features of capitalism attract increasing number of people still living under the socialist system of income earning to jump into the free market. On the other hand, some obvious social guarantees and benefits provided by the old socialist system often prevail in this psychological struggle, and some people are very reluctant to drop out of the system of social care. One can thus expect a variety of different types of individual and social behaviour.

The new specific features and relationships created by the unique process of the socialist/capitalist system transition are a big challenge to economics as a science. One has to describe the observed processes and to introduce new terms and relationships, if necessary. The principal question is - Whether it is possible to describe the transition process in a functional form or whether the process is stochastic and can be described only in statistical terms?

The goal of this study is to develop a model for the transition based on some simple assumptions about the economic state during the transition period and to predict behaviour of some objective and measured economic parameters during the last 15 years and in the future while the transition period has not yet complete. We are here not referring to any other study of the process but concentrate only on the description of the model and results. The only sources used are statistical agencies and databases providing original data on population and GDP.

## 1. Per capita GDP in the former socialist countries 1950-1989

Several Central and East European countries joined the EU in 2004 and some more will join it in 2007. These countries had an almost 40 year history of economic development governed by rules of the socialist system. The countries of the Former Soviet Union (FSU) had an even longer period of socialism reigning in economic life. It is of interest to understand in economic terms what happened during these socialist years and during the period of transformation from socialism to the current state. How far are the countries from the pure capitalist state or do they still bear some elements of socialism so far?

The simplest question one can formulate about the efficiency of the socialist economic system is - What was the average economic growth in these countries in comparison to the most successful capitalist countries? In order to avoid potential geographic effects, only the developed European countries and the USA are used in the comparison. Figure 1 illustrates some relevant results. The original GDP data were obtained from the Groningen Growth and Development Center web-site [3] and represent the PPP estimates. Continuous annual data on real GDP span the period from 1950 to 2003 for the OECD members and some former socialist countries. This comparison is constrained by the continuous observations and does not consider countries with any data gaps. Thus, only the USSR, Poland, Hungary, Czechoslovakia, Bulgaria, Yugoslavia, and Romania are represented.

The former socialist countries (FSC) demonstrated a relatively high rate of per capita real GDP growth from 1950 to 1989. The end of the averaging period is chosen to separate the two principally different periods - the socialist period and the period of the transition to the capitalist system. This is not the exact date, which is derived numerically and used for each of the countries separately, but is a watershed between the two epochs. A more accurate start point of the transition period is determined for each of the studied countries below.

The observed average rate of growth for the FSC lies slightly above the corresponding rates for the most developed countries such as the USA, Switzerland, the UK, etc. On the other hand, the average growth rate in less developed capitalist counties (in terms of per capita GDP) such as Greece, Spain, and Portugal, is higher than in the FSC and in the most developed countries. This observation confirms the concept of the economic growth developed in [2] and briefly discussed below in section 2: the larger the per capita real GDP the lower the economic trend or the growth potential. Hence, the measured rate of economic growth in the FSC was consistently lower than one could expect from the per capita real GDP in these countries from 1950 to 1989.

Figure 2 displays the per capita GDP evolution in the FSC in comparison with the USA and Greece. The FSC had the per capita GDP during the whole period from 1950 to 1989 well below that in the USA in 1950. Greece represents a country which has been developing along the capitalist path with some temporary difficulties. The rate of economic growth in Greece was not high enough even to maintain the absolute lag behind the USA. The estimates of per capita GDP shown in Figure 2 (here and below PPP estimates are used for all the countries except the USA) are used for the modelling of the socialist period and the transition period as well.

By definition, the per capita GDP is determined as the total GDP divided by the total population. Because only people of 15 and above years of age contribute to GDP or earn income, the per capita GDP has to be corrected for the ratio of the population above 15 years of age to the total population. This correction was crucial for the analysis of the GDP evolution in the USA [1]. In many countries, relevant population data are not available, however. So, the per capita estimates made for the total population are used instead. This substitution may result in a somewhat inaccurate prediction of the GDP growth.

The under-performance of the FSC during the socialist era in comparison with the developed capitalist countries is obvious if we consider the relationship between the GDP growth rate and the attained level of GDP per capita. The next question is - What is the maximum theoretical rate of the GDP growth that one could expect in the FSC if they were developing as capitalist countries? - or - How much did they lose during the socialist years?

## 2. Macroeconomic model for the capitalist system evolution

As established in [1], the economic trend is numerically equal to the reciprocal value of $T_{cr}$ - the duration of the period of the mean income growth with increasing work experience. The current (2005) value of $T_{cr}$ in the USA is about 40 years. More specifically, the economic growth rate is defined by the following dependence between $T_{cr}$ and the number of people of some defining age:

$$d(GDP(i))/dt = (GDP(i)-GDP(i-1)/GDP(i-1)=0.5(N(i)-N(i-1))/N(i-1)+1/T_{cr}(i-1) \quad (1)$$

where $d(GDP(i))/dt$ is the real GDP growth rate for the time unit interval $dt=i-(i-1)=1$ (day, month, quarter, year) between the times $i$ and $i-1$, $N(i)$ is the number of people of some defining age and specific for every country at time $i$, and $N(i-1)$ is the same at the previous time $i-1$. Completing the system of equations for the economic growth is the relationship between the per capita real GDP growth rate and $T_{cr}$:

$$T_{cr}(i)=T_{cr}(i-1)\times sqrt[1+\Delta GDP(i))/GDP(i-1)-\Delta NT(i)/NT(i-1)] \quad (2)$$

where $\Delta NT(i)/NT(i-1)$ is the relative change of the total population above 15 years of age, $NT$, during the same period of time. This term is used to show the difference between the real GDP and real GDP per capita. The population distribution is an external parameter for the study, which does not depend (in the framework of the model) on the economic growth.

The first term in equation (1), apparently, oscillates rapidly in time and has a cumulative value for the period from 1950 to 2003 much lower than that of the second term. Thus, one can neglect the first term in the first approximation and assume that the GDP growth rate in the long run depends only on the economic trend, $1/T_{cr}$. Then using the two data sets of real GDP and real GDP per capita one can compare the prediction by (1) and (2) with the actual data of per capita GDP growth rate. Figure 3 presents such a comparison for the USA. The data were obtained from the U.S. Census Bureau and the U.S. Bureau of Economic Analysis official web-sites [4, 5]. The actual values and values predicted by the (1) and (2) dependence of the real GDP per capita are shown.

The original annual data for the per capita GDP growth rate have positive and negative values. In order to carry out a power law regression analysis, the data are shifted up by 0.15, effectively resulting in all positive values. The obtained trend line is shown in Figure 3 by a solid thick line and this is considered further below as predicting the economic trend. The theoretical economic trend for the per capita GDP is very close to the actual trend line. Because of the large positive shift in the original growth rate data the actual trend line has a biased power law exponent.

The actual unbiased trend line for the real GDP growth rate is obtained by averaging of the original growth rates over wide enough moving time windows and conducting the power law regression on these smoothed data. The results obtained by applying this standard procedure are shown in Figure 4. The time window width is 10 years and the time step is 1 year. The trend line has the exponent of - 0.497, which is very close to the theoretical value of -0.5, as one can see from (2). This observation demonstrates the applicability of relationship (1) and (2).

The correction for the population above 15 years of age as discussed above is also applied to the per capita GDP data. The obtained regression and theoretical trend curves are show in Figure 5. The correction improves the consistency of the observed and theoretical curves.

The trend line in Figure 3, when corrected for the real dollar change from 1990 to 1999, is used below to estimate the potential economic trend for the countries in transition, depending on the measured value of per capita GDP in the countries. The exact relationship for the USA is as follows: *per capita GDP growth rate = 63.65x(per capita GDP)*$^{-0.8277}$.

## 3. Modelling of the socialism/capitalism transition process

There is a strict relationship between efforts to increase personal income (and the total income or GDP as a sum of all the personal incomes) and the rate of income growth [1]. The efforts or capabilities to earn money are evenly distributed among the people with age above

15. When a person reaches some critical work experience $T_{cr}$, his/her capability to earn money drops to zero and the personal income rolls off exponentially. The economic development of a socialist system is similar to that of the capitalist system. The only difference is that the personal income in the socialist system has an upper limit and the people with the highest incomes can not contribute more to the total GDP. The overall income deficiency arising from the upper limit on income is about 15% of the GDP. This deficiency relative to the capitalist system resulted in slower economic growth and an increasing lag behind the developed countries from 1950 to 1989.

One can characterize the development of a socialist system using the same general relationship of income growth and capability to earn money as observed for the low-income end of the personal income distribution in the capitalist countries [1,2]. Such an approach results in obtaining numerical values of the defining parameters for the corresponding microeconomic model. When the FSC and the countries of the FSU passed the watershed in 1989-1991 and stepped into the transition zone, the capability to earn money under the socialist system effectively dropped to zero as in the case of an individual reaching the critical age $T_{cr}$ [1]. Hence, the socialist system has undergone a free decay process with the disintegration rate proportional to that of the current level from the level attained by 1989. Such a process is characterized by an exponential roll off. The index of this exponential decrease can be determined from the corresponding observations. From the microeconomic point of view the process of disintegration can be described by a probability, $p$, constant in time, for a person to drop out of the socialist system of payment. Then the number of people leaving the socialist system per year is proportional to the share of the total number of people belonging to the socialist system so far, $Ms(t)$**,** with some coefficient $\alpha$s. This process is analogous top the radioactive decay and is governed by the same exponential law $Ms(t)\sim exp(-a_s t)$.

When the socialist system starts to decay, the capitalist system starts to build up. The process of the capitalist system development consists of two different processes. The first is the increase in the portion of the population governed by the capitalist personal income distribution. The system starts with a zero value because no elements of the capitalist system had existed in the country before the process started. From the microeconomic point of view, the process is described by probability, $q$, constant in time, for a person to enter the capitalist system of personal income earning. This probability is constant for the population and differs from the probability to leave the socialist system, $p$. Moreover, the sum of $p$ and $q$ is not equal to one. The portion of the total population entering the capitalist system per year is proportional with coefficient $\alpha_c$ to the number of people out of the capitalist system so far, $1-M_c(t)$, where $M_c(t)$ is the current portion of the capitalist system. Hence, the evolution of the capitalist portion of population is described by a relationship of the type $M_c(t) \sim (1-exp(-\alpha_c t))$.

Hence, the working age population (everybody above 15 years of age) during the transition period is shared among three reservoirs: socialist, capitalist, and not in either. The first reservoir is decaying at the rate $-\alpha_s exp(-\alpha_s t)$, and the second is growing at the rate $\alpha_c exp(-\alpha_c t)$. In order to keep the systems separate (no person participating in both principal systems) the rate of decay has to be larger than the rate of growth. This requires the inequality $\alpha_s \geq \alpha_c$. This condition effectively creates the third reservoir for the people temporarily being in neither of the two systems. This is not the same as to be unemployed because the latter is built into the capitalist system. It means to be not in either of the main systems of income distribution. For example, it could be a part of a local hidden economy or grey market.

The second growth process of the capitalist system represents the standard economic growth as routinely observed in developed countries. It also starts with a zero value of per capita GDP produced by the first participants of the new capitalist system. At later stages of the development, people enter the system at the current value of per capita GDP. Effectively, it means that newcomers reproduce in their own group the existing system of the personal income distribution. A similar effect is observed in the capitalist system with the younger generations entering the economy [1].

The upper limit of the long term growth rate of the capitalist system (economic trend) is defined by the growth observed in the USA, i.e. the growth rate is controlled by the efforts and dissipation according to the defining relationships as discussed in [1]. In the short term, this process can be described by an exponential growth with some constant index $\alpha$. In the long term relationship (2) indicates a monotonic decrease of $\alpha$ with increasing per capita GDP.

So, the defining equations describing the per capita GDP evolution during the transition from socialism to capitalism are as follows:

$$M(t)=M_s(t)+M_c(t)$$

$$M_s(t)=M^0{}_s exp(-\alpha_s t) \qquad (3)$$

$$M_c(t)=(1-M^0{}_c)(1-exp(-\alpha_c t))exp(\alpha t)$$

where $M(t)$ is the total per capita GDP at time $t$, $M_s(t)$ and $M_c(t)$ are the portions of the socialist and capitalist systems in the total per capita GDP, $M(t)$, $M^0{}_s$ *and* $M^0{}_c$ are the initial portions of the socialist and capitalist system in the total per capita GDP, $M(0)$ **(**we assume $M^0{}_c=0$, $M^0{}_s=1$**),** $\alpha_s$ is the dissipation factor of the socialist system decay, $\alpha_c$ is the dissipation factor of the capitalist system growth, and $\alpha$ is the observed economic trend of the capitalist system growth. In (3) we implicitly assume that the portions of the socialist and capitalist systems in the total per capita GDP are proportional to the share of the system in the total population.

We neglect the term for internal growth (positive or negative) of the socialist system in equation (3) because of the fast decay of the system itself. In any case, the growth of the socialist system is also an exponential process with some index $\alpha_{si}$. This index can be effectively included in the index $\alpha_s$.

The transition process between the socialist and capitalist systems is described by an equation containing three defining parameters. The exact values of the parameters can be determined by a standard matching process: the parameters are varied in the reasonable range to get the best fit between the observed and predicted time histories of $M(t)$. We did not apply any formal statistical procedure like $\chi$-squared criteria, but only used the "eye-fit", which usually provides reasonably good accuracy. Our principal goal is not to find the exact values of the defining parameters, but to demonstrate the consistency of the modelling approach.

Figures 6 and 7 illustrate the procedure as applied to Russia. The starting point of the transition process is 1991, when the largest republics of the Former Soviet Union announced their independence and initiated the new economic policy towards capitalism. The Groningen data set provides relevant values of the per capita GDP (Geary-Khamis PPP) for the studied period. The initial presence of the socialist system, $M^0{}_s$, is 1.0, i.e. no capitalist elements before 1991 had been permitted in Russia. The best-fit model is characterised by the following defining parameters: $\alpha_s$=0.24, $\alpha_c$=0.066, $\alpha$=0.033. The socialist system portion in the total GDP per capita drops relatively slowly compared to the Central European countries, as is shown below. The current (2005) value of the socialist system portion is only 2%. There were large elements of socialism in the mid 90s, however, when the sharpest economic decrease was observed.

The build-up of the capitalist system is also characterised by a relatively slow rate. Parameter $\alpha_c$=0.066 provides only about 65% of the population to be in the capitalist system 15 years after the start. The residual 33% of the total population are not then formally included in either of the main income distribution systems. This reservoir is potentially feeding the black market and antisocial groups. Moreover, the growth rate of the capitalist system portion in Russia is currently low and special efforts are needed to create conditions to enable more people to enter the capitalist system of production and distribution. The 33% portion of "non-attributed" population is one of the highest among the former socialist countries.

Figure 7 displays the evolution of the predicted and measured per capita GDP in Russia. The observed coincidence is very good given the simplicity of the model and possible errors in the PPP per capita GDP estimates. In general, only relatively smooth and coordinated changes have occurred in the Russian economy since 1991. The transition process has two branches: one downward (from 1991 to 1998) and one upward (from 1998 to present). Currently, the economic growth in term of the per capita GDP is fuelled by two processes: the internal growth of about 3.3% and the growth of the capitalist system portion of about 2.2%, making about 7.3% (=0.022x0.033) in total in 2004. Both factors of growth are decaying in time. The first one is limited by the attained level of the per capita GDP. The growth rate limit corresponding to the initial Russian per capita GDP of \$7371 in 1991 was 4.0% (=63.65x7371$^{-0.8277}$), and the current value of the limit is also about 4.1%. In relationship (3), we neglect the potential growth rate change during the transition period.

The value of the current internal growth of the per capita GDP in Russia is somewhat lower than its potential value as defined by the growth rate observed in the USA at the same level of the per capita GDP. The difference is not large, however, and may be partially explained by the accuracy of the PPP per capita GDP and GDP growth rate estimates. The Russian Federation has some room to accelerate the growth both by increasing the portion of the capitalist system and by using the whole potential of the capitalist system growth.

Figure 8 displays the observed and predicted evolution of the per capita GDP in Hungary. The start point for process is 1990. The observed GDP changes are not as smooth as those in Russia. The deviations from the predicted curve may be due to the effects of the population changes defined by the first term in the equation (1). These deviations are only short period oscillations and do not change the economic trend. Currently, the observed and predicted curves are in a very good agreement and one can expect the future GDP per capita developing along the predicted line. The evolution of the socialist and capitalist shares in Hungary is shown in Figure 9. Because the defining parameters, $\alpha_s$=0.40 and $\alpha_c$=0.23, are larger than those for Russia, the evolution of the capitalist and socialist portions is much faster. There is no socialist system in Hungary any more and its portion is well below 1%. The capitalist system has occupied almost the whole available model space - population portion. The observed growth of the Hungarian economy is totally due to the inherent growth of the capitalist system. The long-term trend of the per capita GDP growth derived by the modelling is 2% ($\alpha$=0.02). This value is below the potential of 4.2% as obtained for the per capita GDP of $6903 in 1989. At present, the theoretical economic trend is 3.7% for the per capita GDP value of $7930. Hence, the Hungarian economy grows at a slower pace than expected.

The Polish GDP per capita evolution is presented in Figure 10. The predicted curve excellently describes the initial part of the transition process from 1989 to 2000. Only minor deviations were observed for this period. In 2000, a major deviation from the predicted curve coincided with two years of very slow growth. The observed economic trend apparently returns to the predicted slope after 2002: the theoretical curve after 2002 is almost parallel to the observed one but lies higher. The deviation may be of the same nature as was discussed for Hungary - high-frequency oscillation due to corresponding change in population of some specific age. Thus one can expect the observed curve to return to the predicted one in some short time in a rebound process. Some irreversible process is also possible. Then the curves will stay parallel in future until the next oscillation - negative or positive.

The observed Polish economic trend is higher than in Hungary - 3%. The potential value for the per capita GDP of $5700 in 1989 is 5.0% and 3.7% for $8000 in 2004. So, currently

Poland uses a large part of the potential economic growth. The portion of the capitalist system in Poland is approaching 100% as in almost all other Central European countries. (Slovenia is an exception.) This large portion has been achieved due to a high value of the index $\alpha_c$=0.23.

The Czech Republic and Slovakia are very close in their defining parameters and evolution shown in Figure 11 and 12. Slovakia has slightly lower per capita GDP and a slightly higher growth rate. The Czech Republic had a slow down period from 1995 to 1997, but then returned to the monotonic growth with $\alpha$=0.017, as represented by the parallel theoretical and observed curves since 1998. The Czech growth is lower than the expected value of 3.1% for the per capita GDP of $10,000 in 2004. In 1990, the expected growth rate was 3.4%.

Slovakia also had a short period of deceleration in 1998 and 1999. Since then it has been consistently returning to the predicted curve with $\alpha$=0.019. The theoretical trend of the per capita GDP growth rate in Slovakia has changed from 3.8% in 1990 to 3.4% in 2004.

Two more examples are Bulgaria and Slovenia. They are characterised by the smallest and largest among the studied countries per capita GDPs with $6200 and $14632 respectively. The expected growth rates are 4.6% and 2.2%. The actual values are 2.9% for Bulgaria and 2.0% for Slovenia. Thus, Bulgaria struggles to use all the advantages of capitalist growth, and Slovenia nearly reaches its highest potential performance. It is interesting that the two countries are characterised by almost the same defining parameters except for the growth rate. These parameters indicate a relatively low rate of the socialist system dismantling and relatively slow implementation of the capitalist system.

The lowest point of the per capita GDP decline for all the Central European countries is at 0.8 of the initial value. This results from the coordinated decay and build-up with an $\alpha_s/\alpha_c$ ratio of 2. The closer is the ratio to 1.0, the shallower the trough in per capita GDP. The ideal case is with the ratio equal to 1.0, when the total population immediately jumps from socialism to capitalism with no delay in the third reservoir. Russia is characterised by a high ratio that is above 3. This makes the trough deep and the transition process long with many people not in either of the main economic systems. Russia does not have the highest ratio among the countries in transition. Moldova has a ratio of 10. This and some other cases are illustrated in Appendix A.

## 4. **The future for some new EU members**

The above discussion results in a conclusion that the transition process has effectively finished in the Central European countries. Thus, the potential rate of per capita GDP growth in the countries is limited by the value defined by the attained level of the GDP per capita, as it is in other developed countries. Any deviations from the theoretical rate can be explained only by variations of single year population of some defining age or by inefficient organization of the economy. The former can only be of a short-term nature. The latter is expressed in the fact that many developed countries are characterized by lower values of economic trend than the USA at the same level of GDP per capita.

Here we assume that the economic growth in the FSC in future will follow its highest potential value depending on the per capita GDP. No inefficiency in the organization of the economy is allowed and the best case scenario for the development is realized in the next 50 years. Equations (1) and (2) completely define the future of the FSC with the start point in 2005. The observed values of the per capita GDP in 2004 are used as initial values.

Figures 15 and 16 show the evolution of the per capita GDP in Hungary and Poland in comparison with the USA. Czech Republic, Slovakia, Bulgaria and Slovenia have histories of the economic development similar to those of Hungary and Poland but with different initial values. The GDP per capita curves are theoretical after 2004 and the observed ones before 2004. The USA curve is much higher and has no flat and downward parts as was observed in Hungary and Poland for a long time. In Poland, the stagnation process started in late 70s. Hungary is characterized by a shorter period of low performance.

The best case scenario results are realized by maintaining for absolute lag behind the USA in terms of GDP per capita. At present, this lag represents about $20,000 dollars (chained 1999). In relative terms, the lag will decrease as a function $X/(X+A)$, where $X$ is the per capita GDP in a given FSC and A is the constant absolute lag in 2004.

## 5. **Discussion and conclusions**

At first glance, the above consideration and the results obtained are similar to many other investigations which sometimes use very sophisticated interpolation and extrapolation procedures in order to describe the observed economic time series. There are several aspects, however, which are inherently different between this study and other interpolations. It was found that the exponential functions not only describe economic processes well, but also result from basic economic relationships at the micro level, which are solutions of the equations defining the personal income distribution and its evolution. One can consider the functions as basic functions or eigenfunctions of the economic processes. Hence, the interpolation we have carried out is based on a full system of eigenfunctions, which potentially can provide the exact description of the observed processes. The assumption of the even distribution of the probability for a person to leave the socialist system or to enter the capitalist system is not only reasonable but also is the simplest possible and consistent with the even distribution of the capability to earn money and the means used to earn money discussed in [1]. As a result, one has a simple system that excellently describes the transition from the socialist system to the capitalist system expressed in terms of per capita GDP.

The per capita GDP is a fundamental parameter defining the economic state in any given country, i.e. the attained level of development. The total GDP depends on the total population but is often misinterpreted to represent the level of economic development of some countries. The mistake is revealed by the fact that the per capita GDP is the only parameter defining the potential rate of economic growth or economic trend. So, a high growth rate is not large achievement for countries with low per capita GDP because it is expected. And some slow growth in developed countries is not a characteristic of underperformance. The countries have to be evaluated by their performance relative to the theoretical value. The best criterion is not the relative per capita GDP growth, but the absolute increase, as is clear from Figures 15 and 16, where the theoretical absolute lag is constant.

As we have seen above, not many of the FSC and FSU countries follow their potential path of growth. The majority find themselves below the potential and only a few have a short term growth rate above the potential (see Appendix A). In the long run, there is an upper limit, however.

The goal of this study is only to estimate exactly the current state of various economies in transition in terms of the model developed, and not to give any specific recommendations on how to improve the situations. The current situation derived from the model provides a good indication to the measures needed to be taken to move the situation in a “positive” direction.

A comparative analysis of various countries in transition is also helpful for understanding what direction of development is positive in terms of per capita GDP.

The principal result of this study is the model for the transition from the socialist to capitalist system. The model is defined by only three parameters, which have very clear analogies in natural sciences. Hence, the interpretation of the obtained results can also be simple. Although the model is easy to use and to interpret, the values of the defining parameters for a particular country could not be obtained beforehand. Every country has a specific set of conditions and internal population relationships. The conditions define the values of the indices. It is hard to judge whether it is possible to change the indices for the countries still in the transition, but any reasonable behaviour has to be directed towards decreasing the ratio $\alpha_s/\alpha_c$ and increasing $\alpha$, if possible.

**Acknowledgments**

The author is grateful to Dr. Wayne Richardson for his constant interest in this study, help and assistance, and fruitful discussions. The manuscript was greatly improved by his critical review.

# Figures

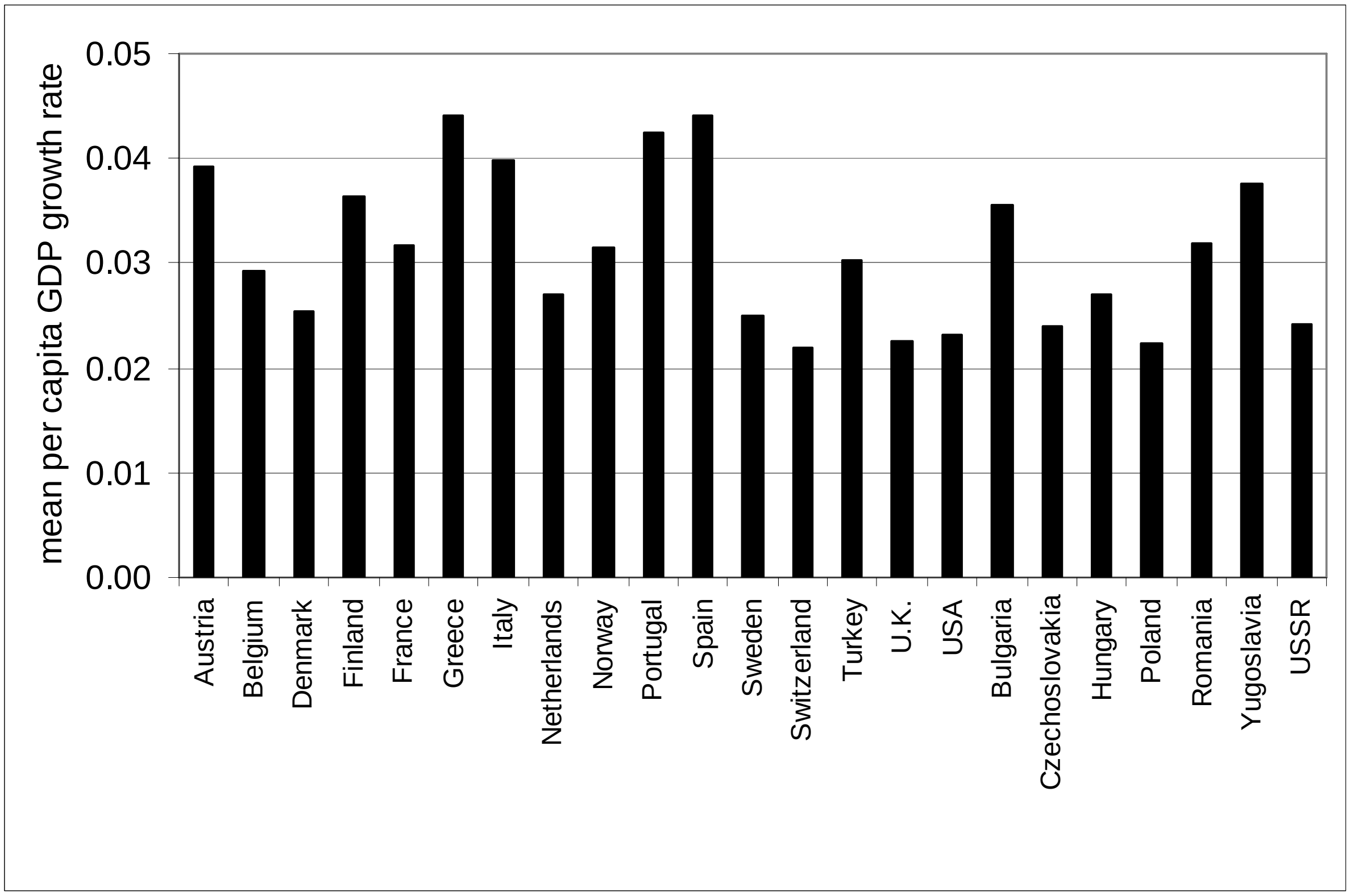

Fig. 1. Average per capita GDP growth rate for some developed counties and the FSC for the period from 1950 to 1989. Greece, Portugal, and Spain have the highest average growth rate. The FSCs match the level of the countries with the highest GDP per capita.

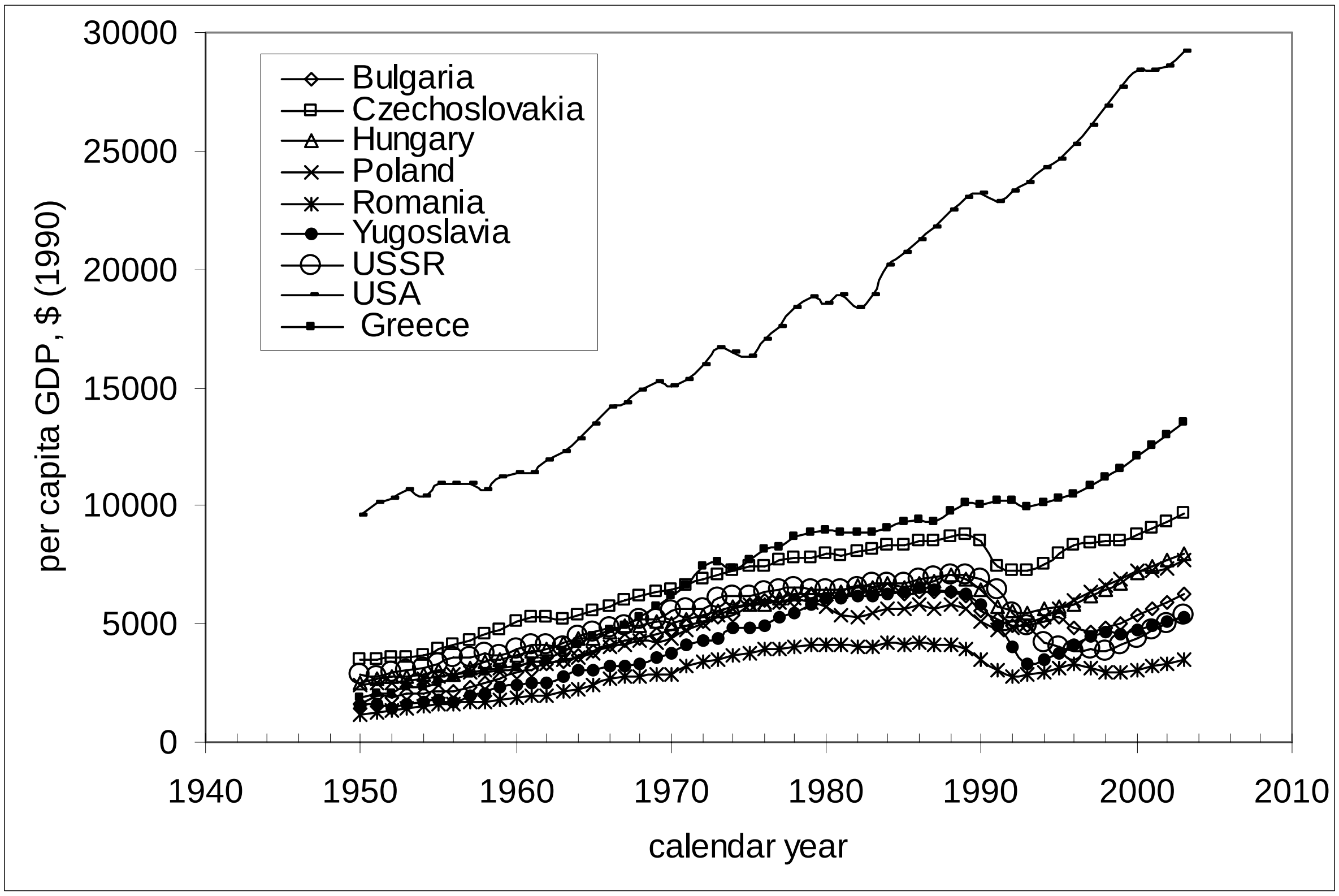

Fig. 2. Evolution of the per capita GDP in the FSC in comparison with the USA and Greece for the period from 1950 to 1989. The absolute gap between the USA and other countries increases with time demonstrating a more efficient economy in the USA.

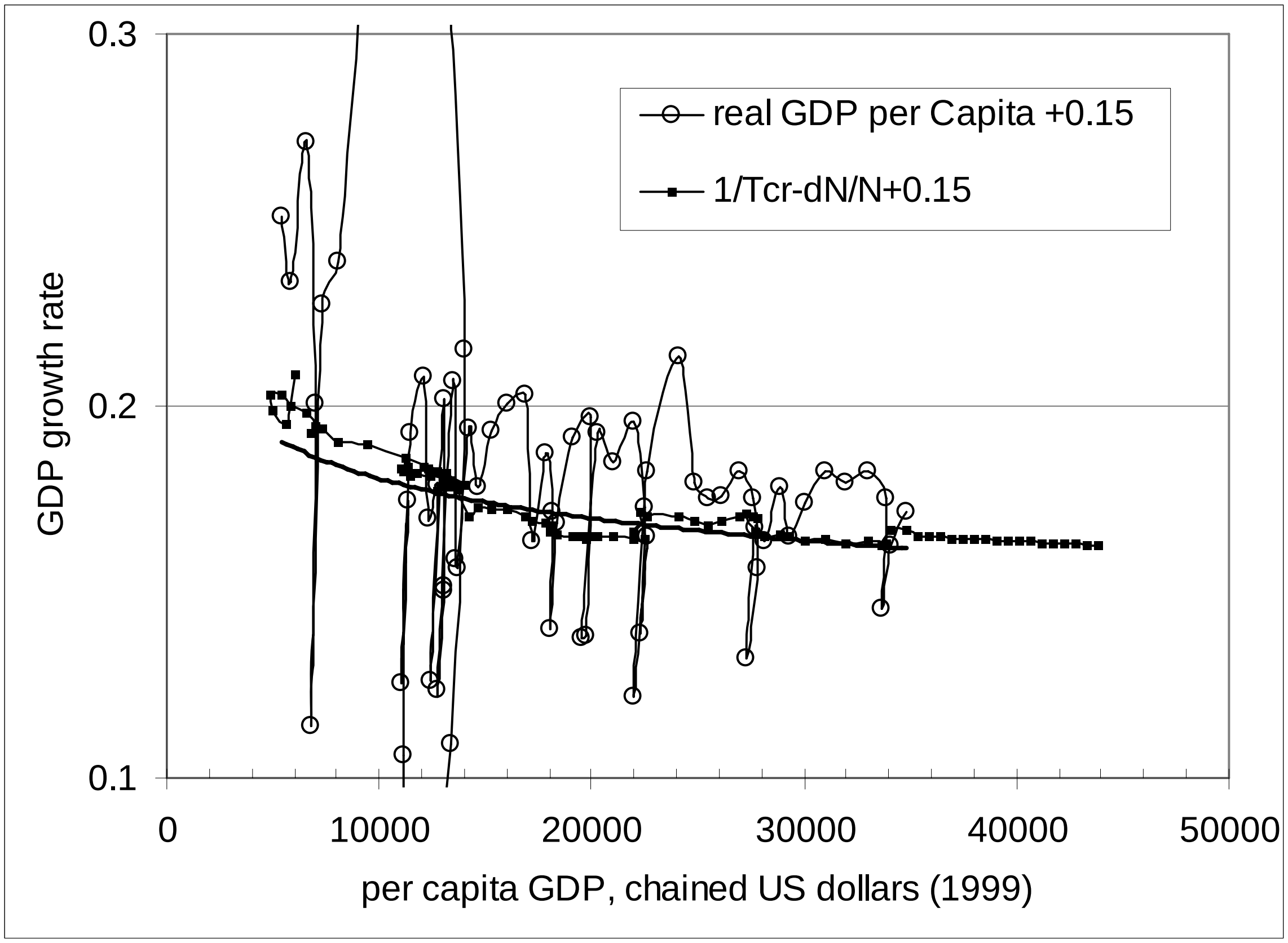

Fig. 3. Real GDP per capita growth rate (chained 1999 US dollars) in the USA from 1930 ($6301) to 2003 ($34831). The measured growth rate values are shifted up by 0.15 in order to make them all positive and to carry out a power law interpolation shown by a solid line. The power law regression line of the observed growth rates (e.g. economic trend) is compared with the theoretical economic trend represented by the relationship 1/Tcr-dN/N+0.15.

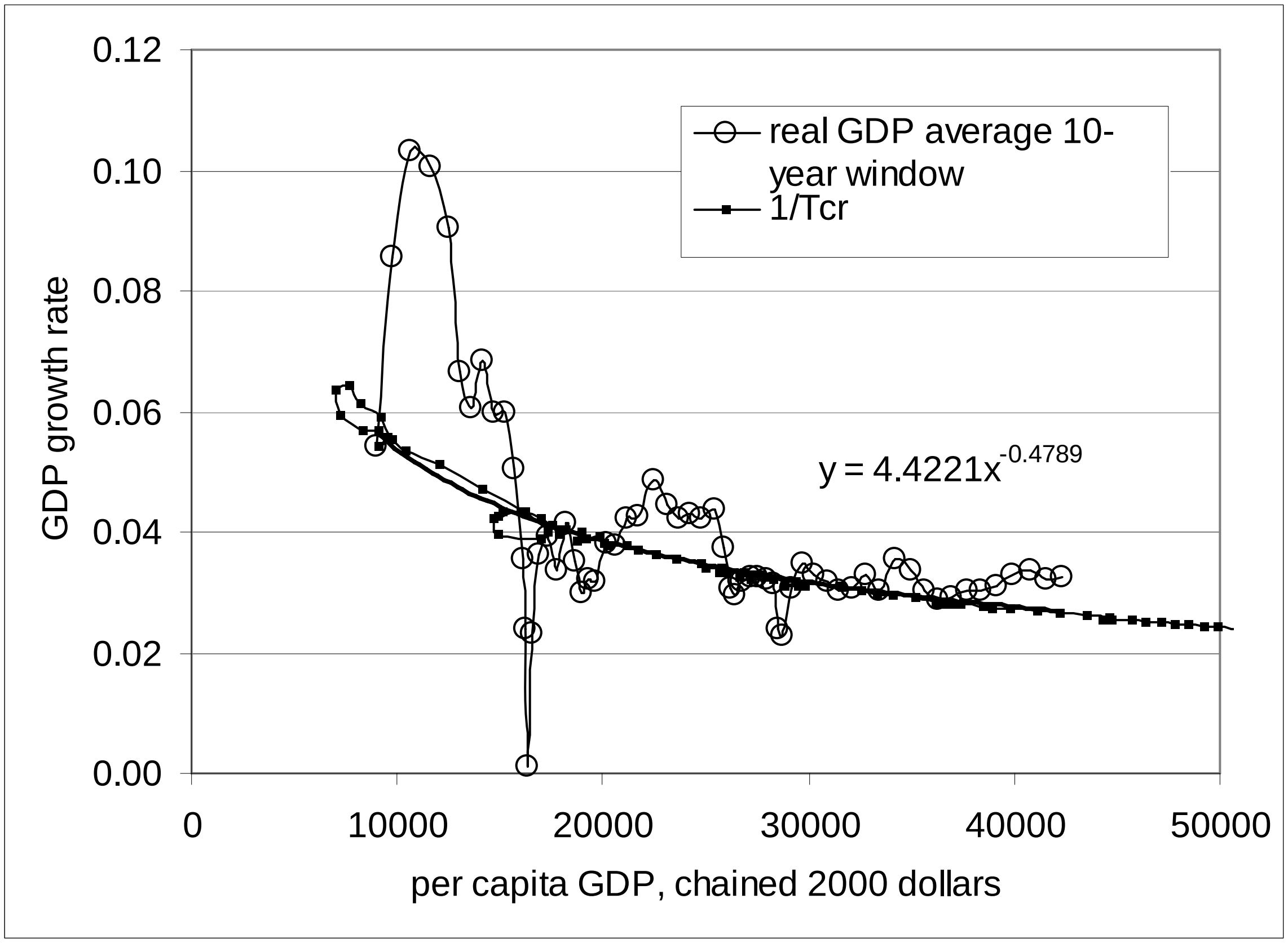

Fig. 4. Power law regression (solid line) of the real GDP growth rate averaged values. The growth rates are smoothed (averaged) using a 10 year wide time window with a 1 year step. The power law exponent is -0.48, i.e. very close to its theoretical value of -0.5. The theoretical curve (1/Tcr) calculated as a square root of the real per capita GDP growth rate is shown for a comparison.

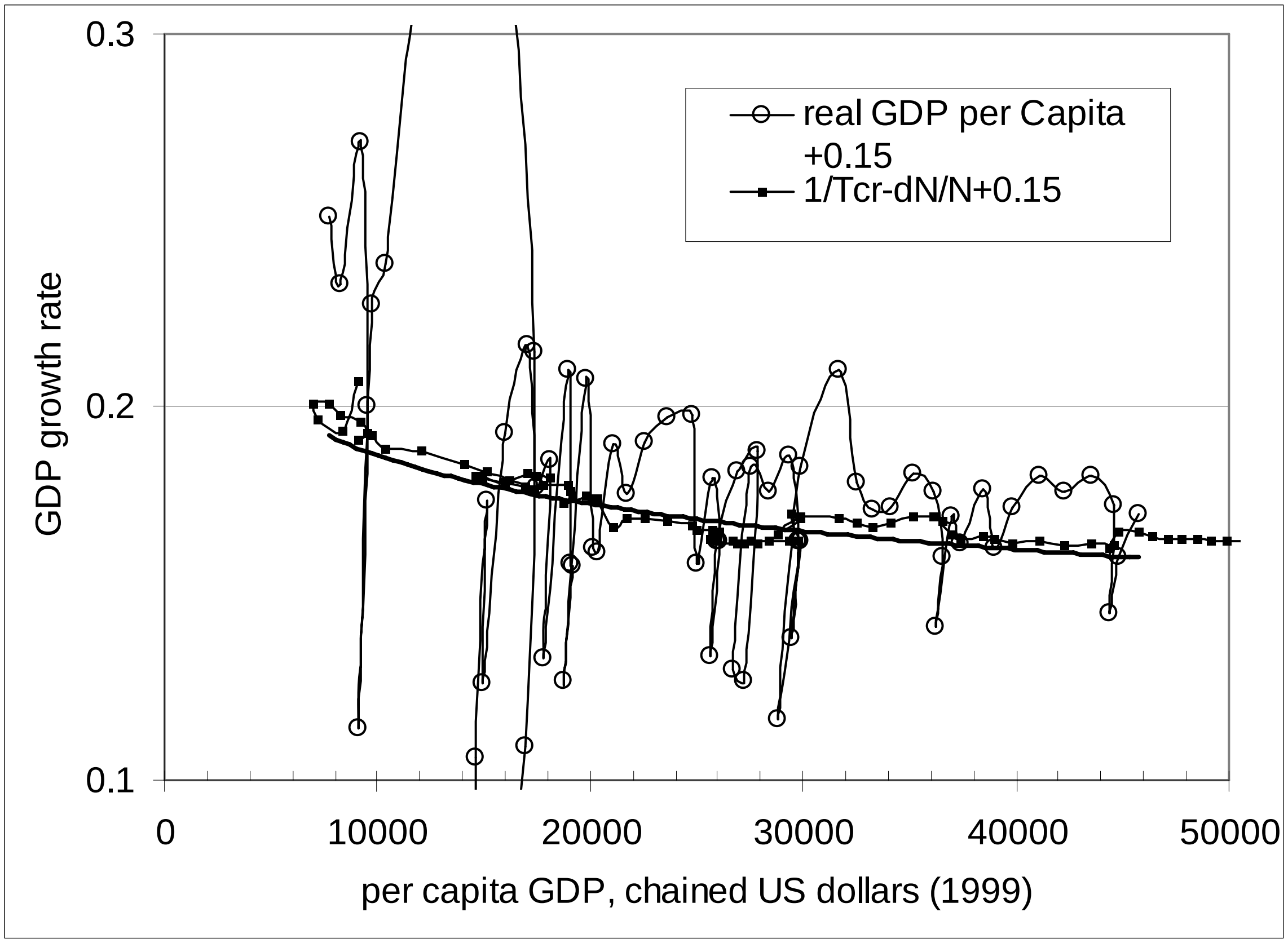

Fig. 5. Real GDP per capita growth rate (chained 1999 US dollars) in the USA as a function of the per capita GDP from 1930 ($9073) to 2003 ($45726). All the GDP growth rate values are shifted up by 0.15 in order to make them all positive and to carry out a power law interpolation shown by a solid line. The GDP per capita values are corrected for the ratio of the total US population and the population above 15 years of age. The power law regression line of the observed growth rates (e.g. economic trend) is compared with the theoretical economic trend represented by the relationship 1/Tcr-dN/N+0.15.

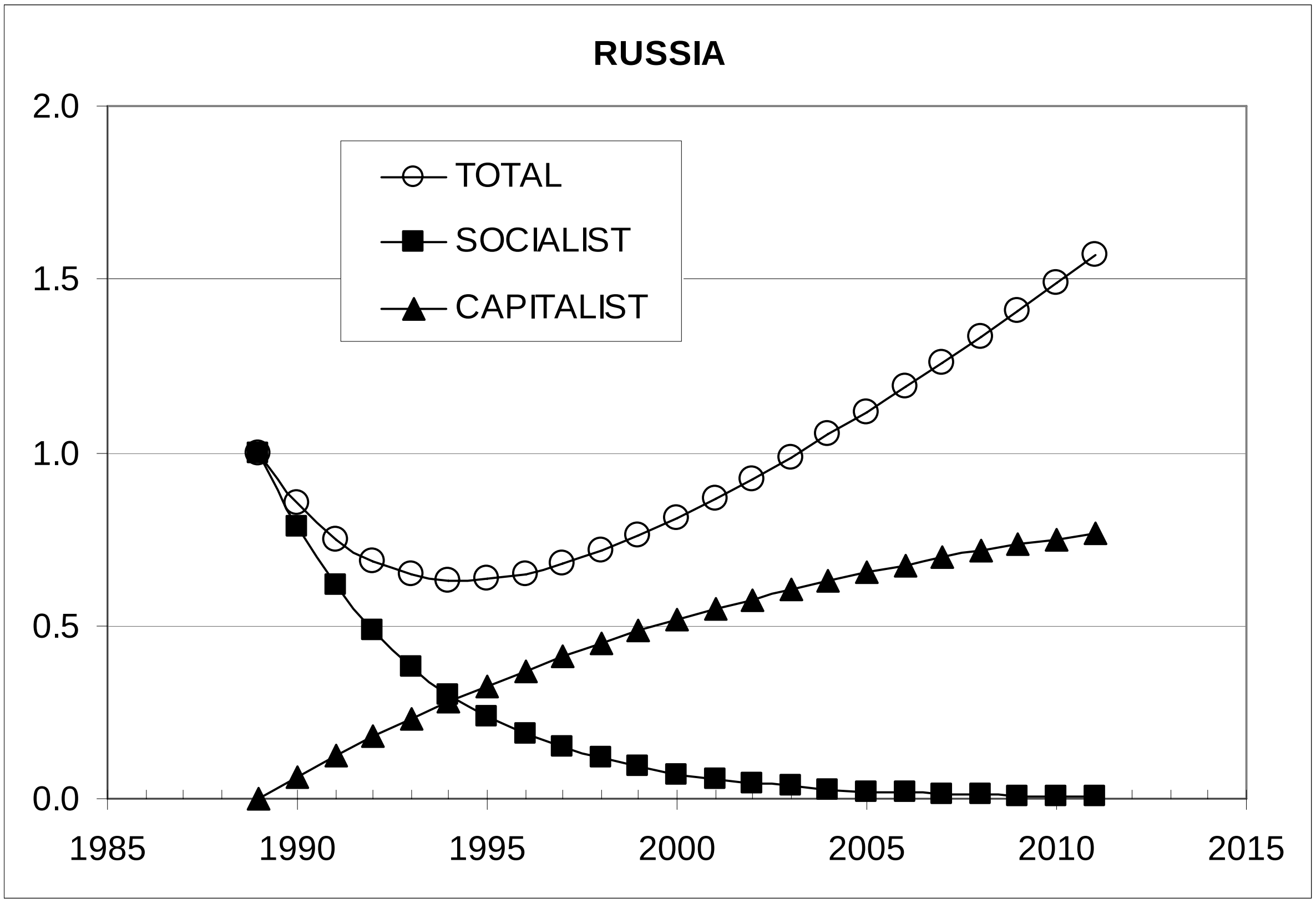

Fig. 6. Free decay of the socialist system in Russia starting in 1991. The rate of disintegration is proportional to the attained level. The best fit model parameters are $\alpha_s$=0.24, $\alpha_c$=0.066 and $\alpha$=0.033. At present, approximately 2% of the socialist system still exists. The capitalist system in Russia grows from zero point in 1991. The portion of the capitalist system is currently about 65%. Thus, about 33% of the population not either in socialist or capitalist system. The total growth of the per capita GDP (TOTAL) in Russia relative to 1991 is defined by the combined evolution of the two elements.

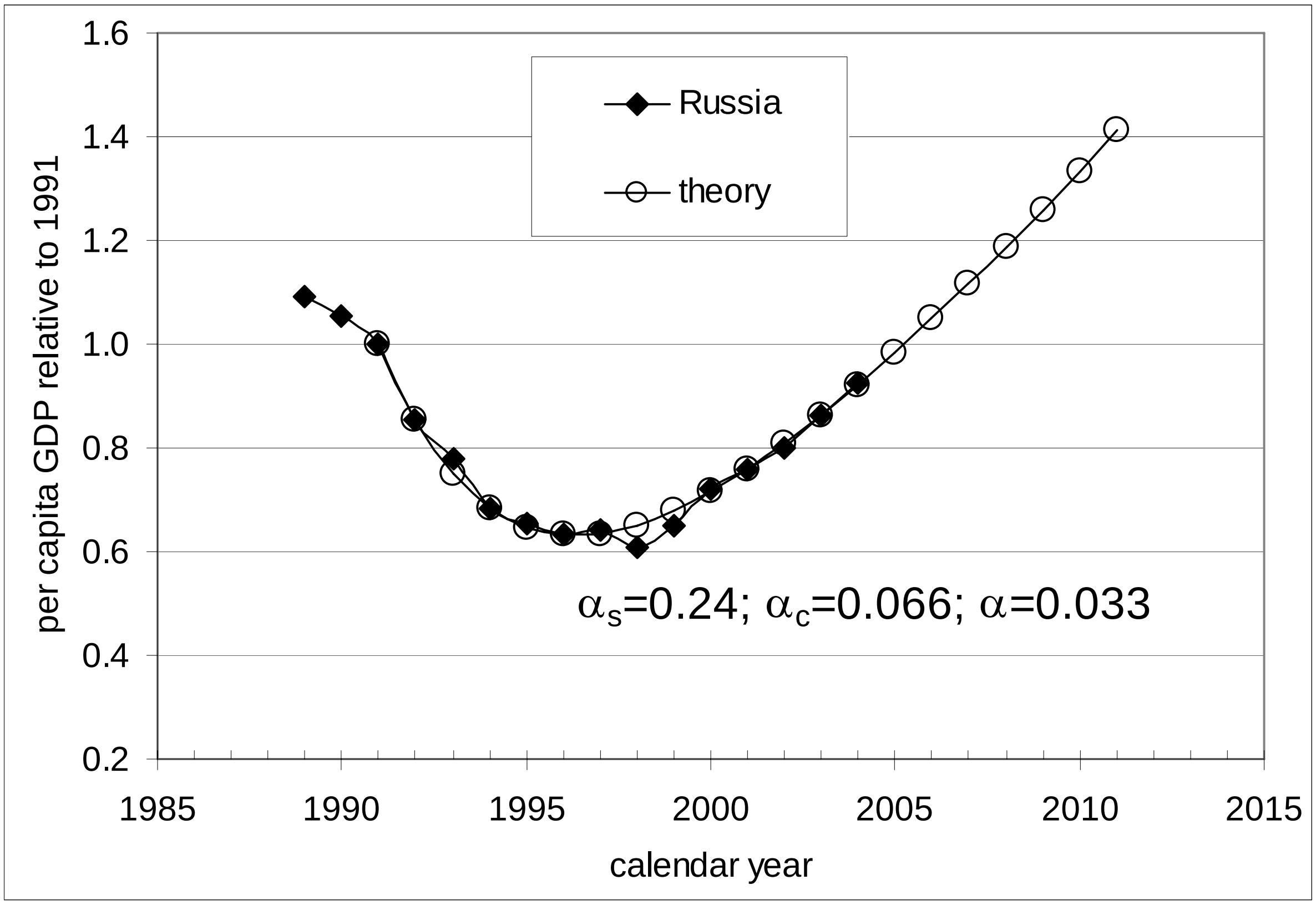

Fig. 7. Comparison of the observed and predicted transition process for the replacement of the socialist system with the capitalist system in Russia. Parameters of the best fit model are indicated.

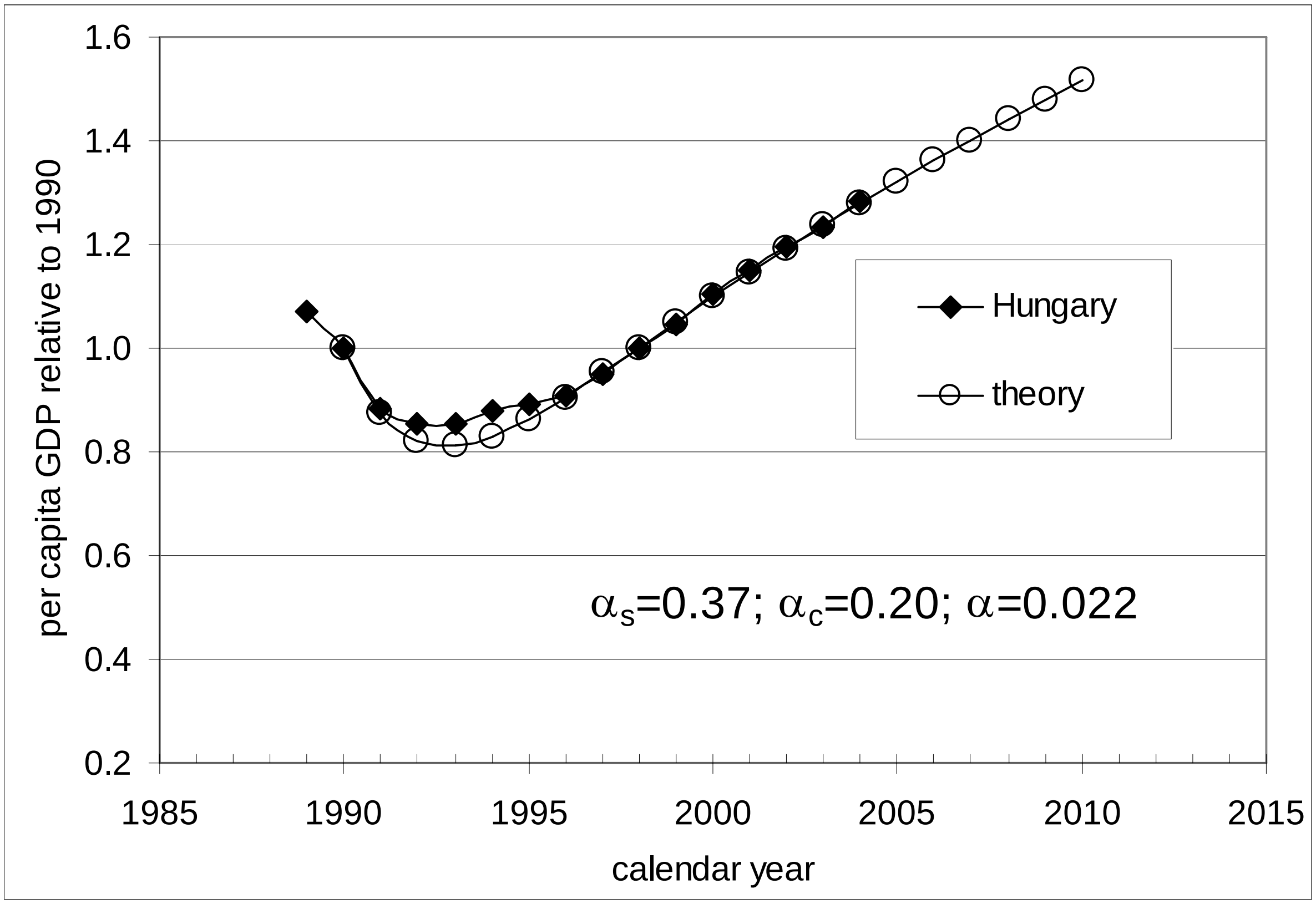

Fig. 8. Comparison of the observed and predicted transition process for the replacement of the socialist system with the capitalist system in Hungary. Parameters of the model are indicated. The per capita GDP grows in Hungary at long-term rate of 2.1%.

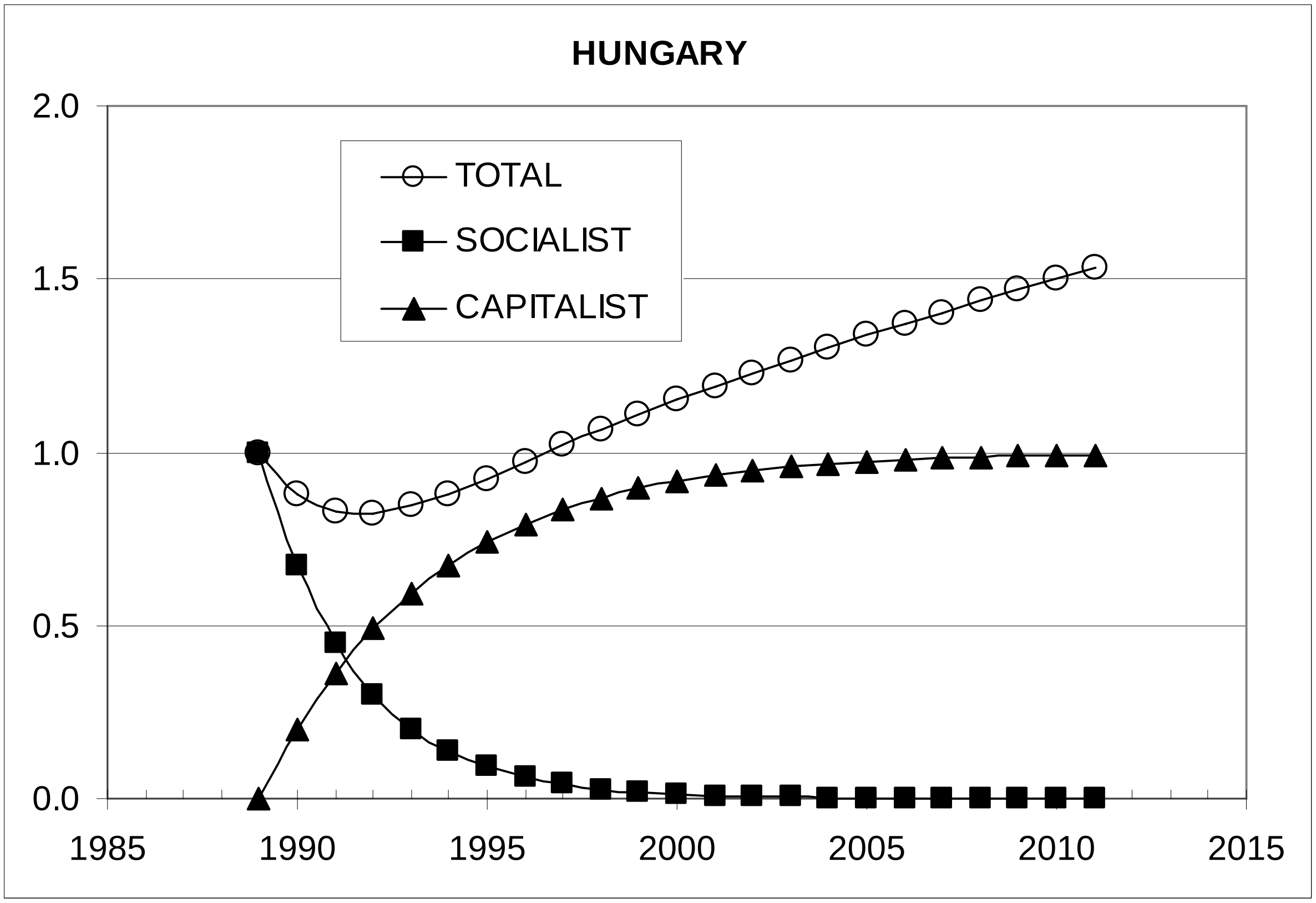

Fig. 9. Free decay of the socialist system in Hungary starting 1989. The best fit defining parameters are $\alpha_s$=0.40, $\alpha_c$=0.22 and $\alpha$=0.021. There is effectively no socialism in Hungary any more. The capitalist system has filled almost the whole available model - population space. The economic growth currently observed is due to the inherent development of the capitalist system.

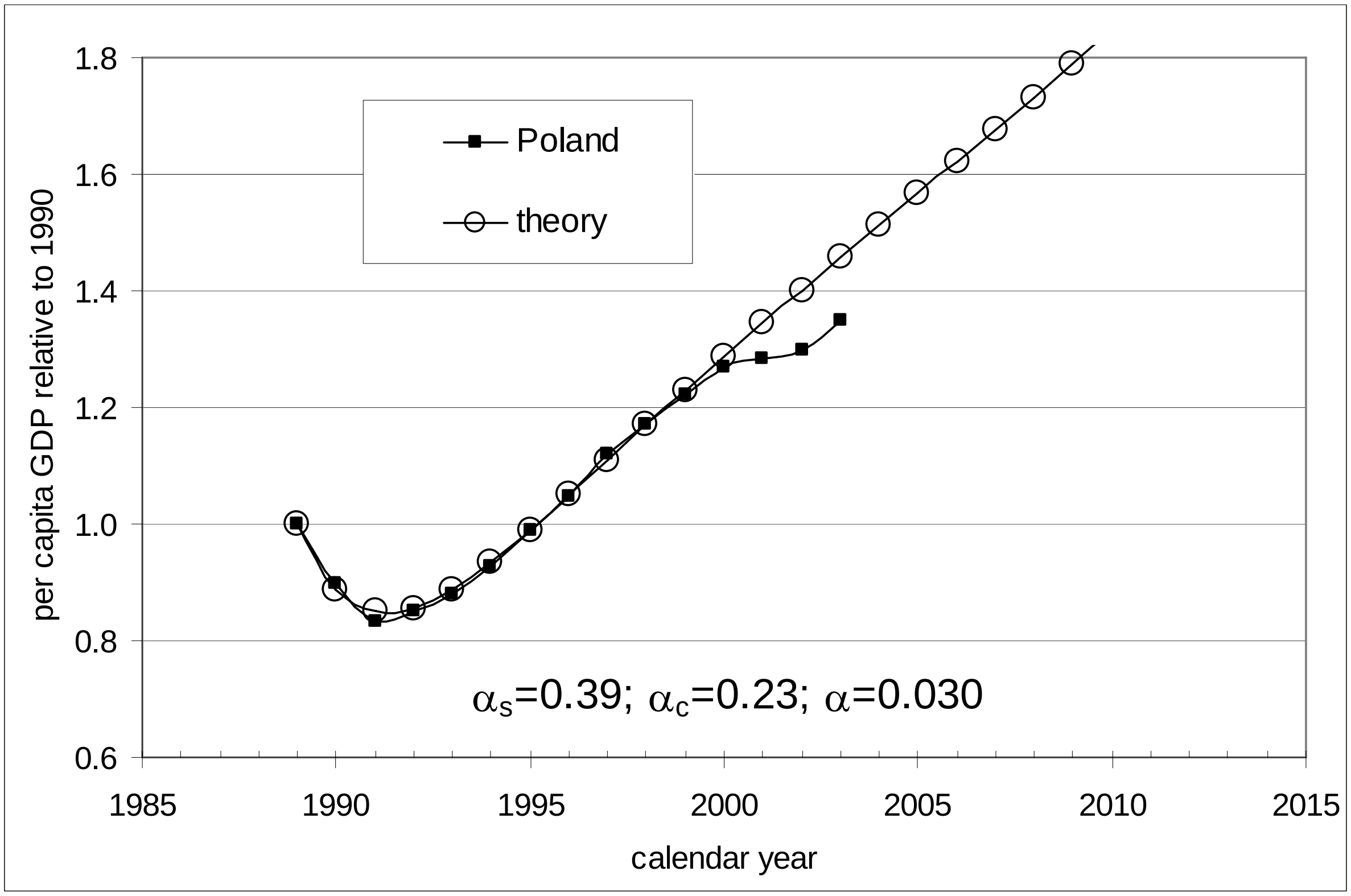

Fig. 10. Comparison of the observed and predicted transition process for the replacement for the socialist system with the capitalist system in Poland. Parameters of the model are indicated. The start point is 1989. Two years of a sluggish economy were observed in 2000 and 2001.

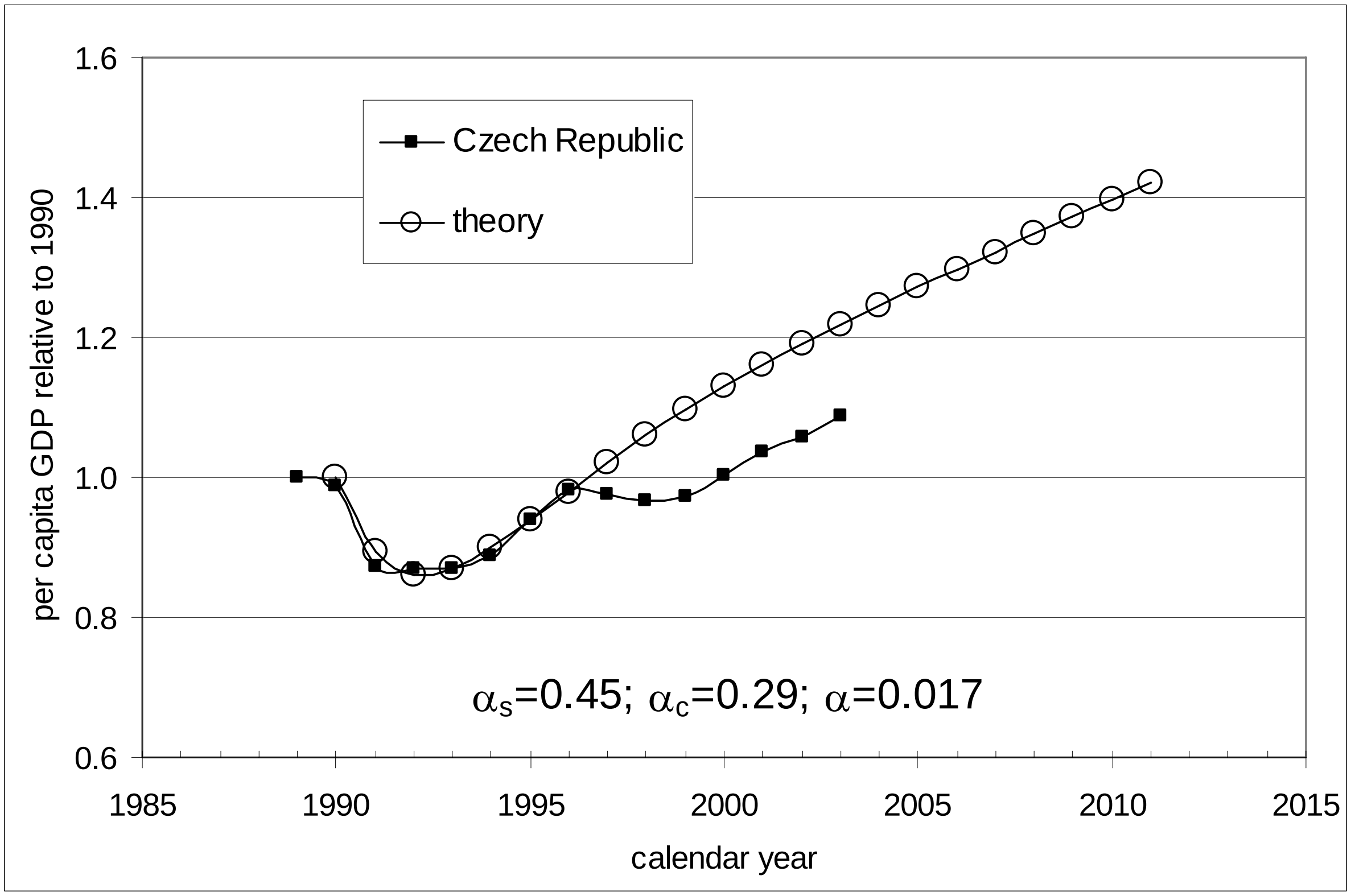

Fig. 11. Comparison of the observed and predicted transition process for the replacement of the socialist system with the capitalist system in the Czech Republic. Parameters of the model are indicated. The start point is 1990. Poor performance lasted for three years - from 1997 to 1999.

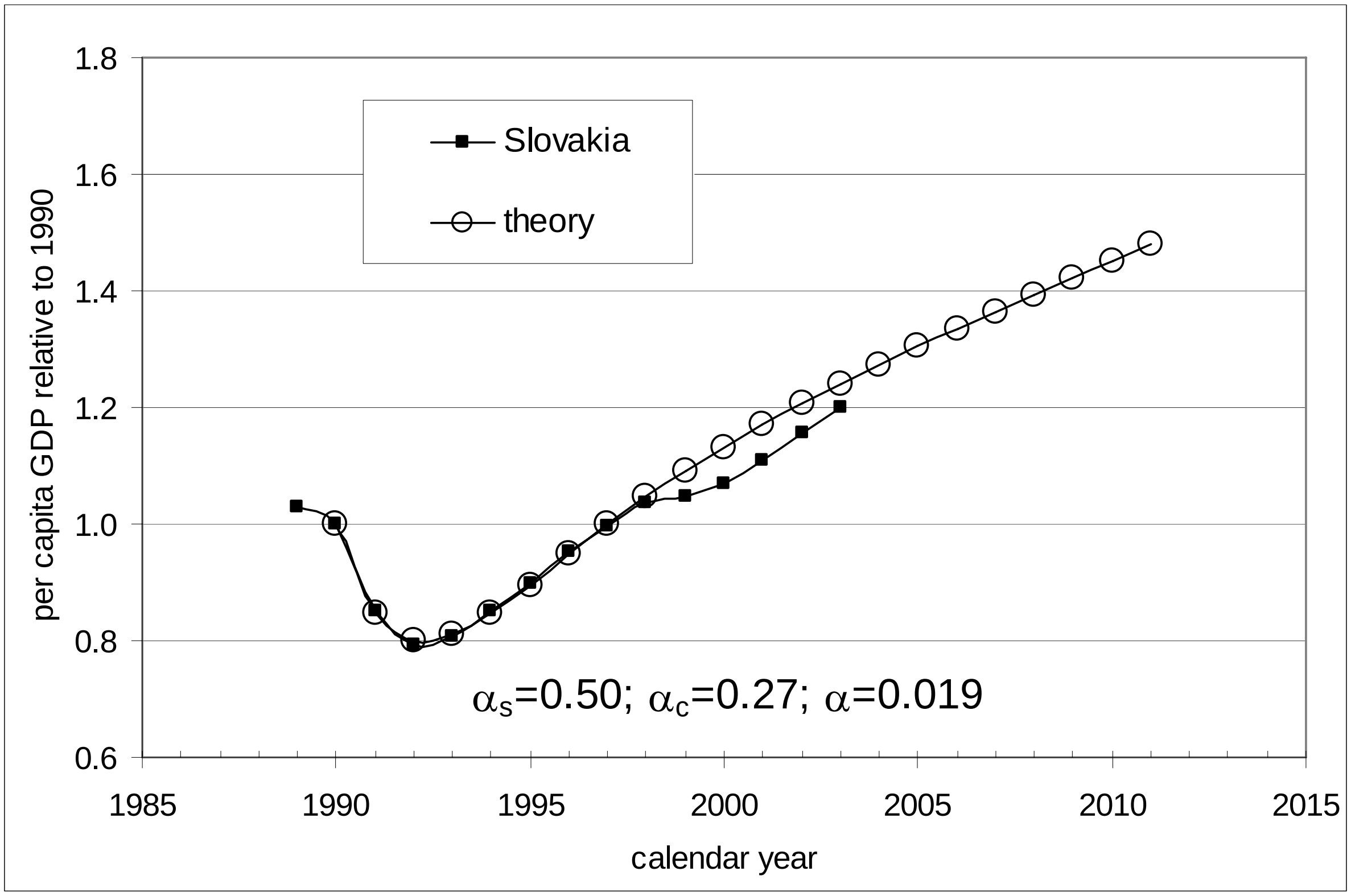

Fig. 12. Comparison of the observed and predicted transition process for the replacement of the socialist system with the capitalist system in Slovakia. Parameters of the model are indicated. The start point is 1990. One year of slow down in 1999 has been almost recovered until now.

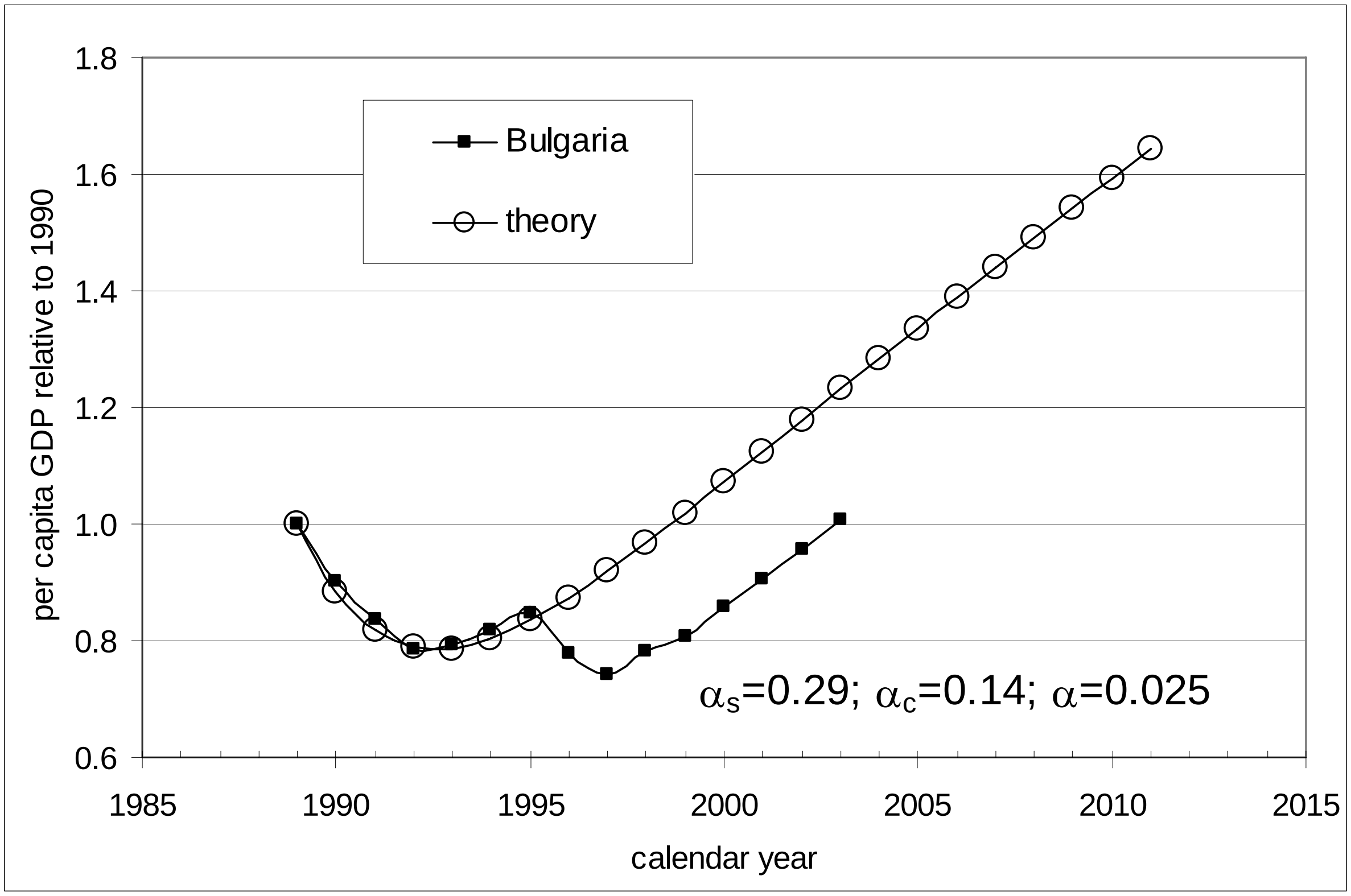

Fig. 13. Comparison of the observed and predicted transition process for the replacement of the socialist system with the capitalist system in Bulgaria. Parameters of the model are indicated. The start point is 1989. Two really poor years 1996 and 1997 delayed growth. The current growth rate is relatively high.

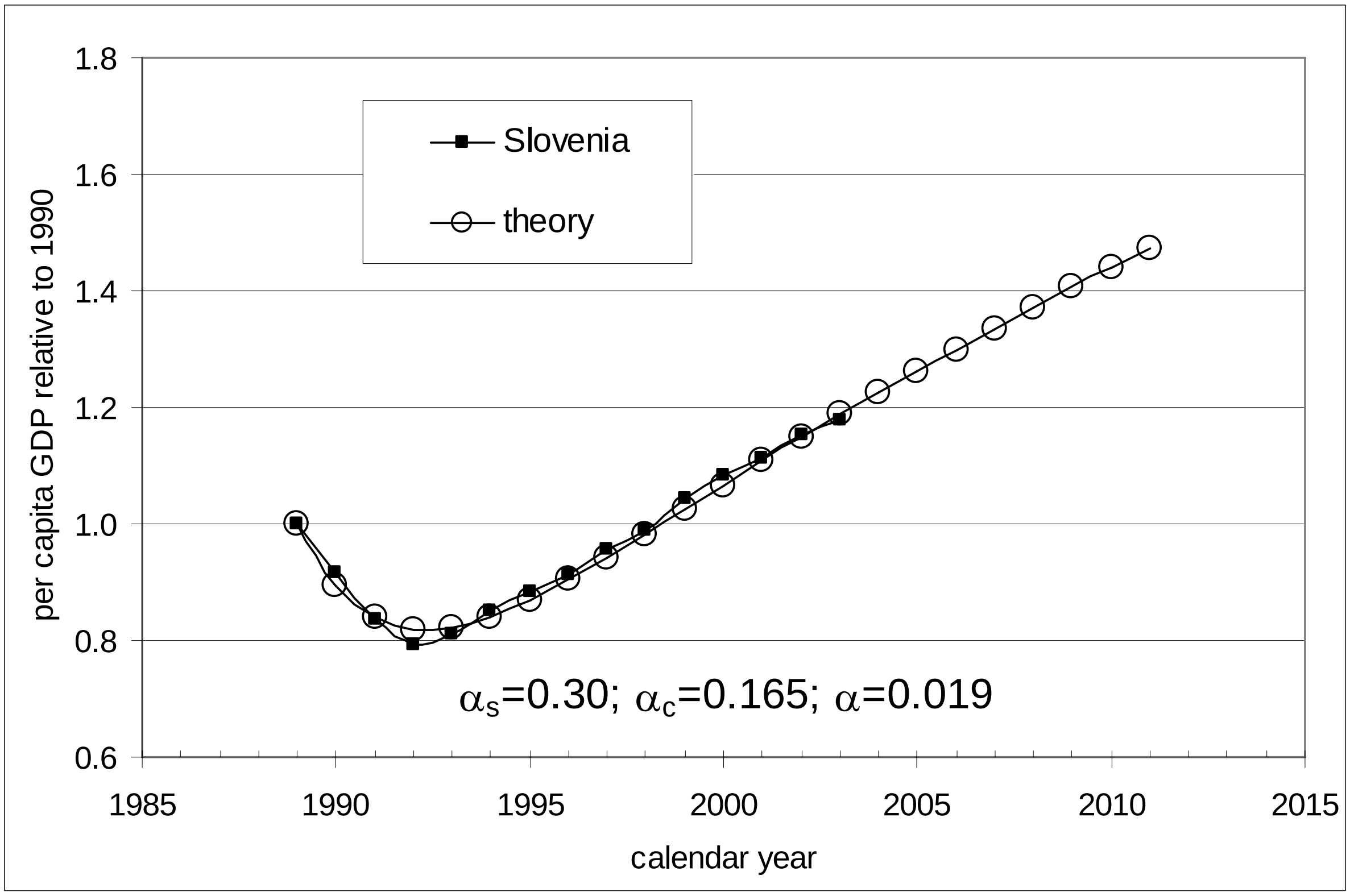

Fig. 14. Comparison of the observed and predicted transition process for the replacement of the socialist system with the capitalist system in Slovenia. Parameters of the model are indicated. The start point is 1989. Only minor divergences from the predicted curve have occurred.

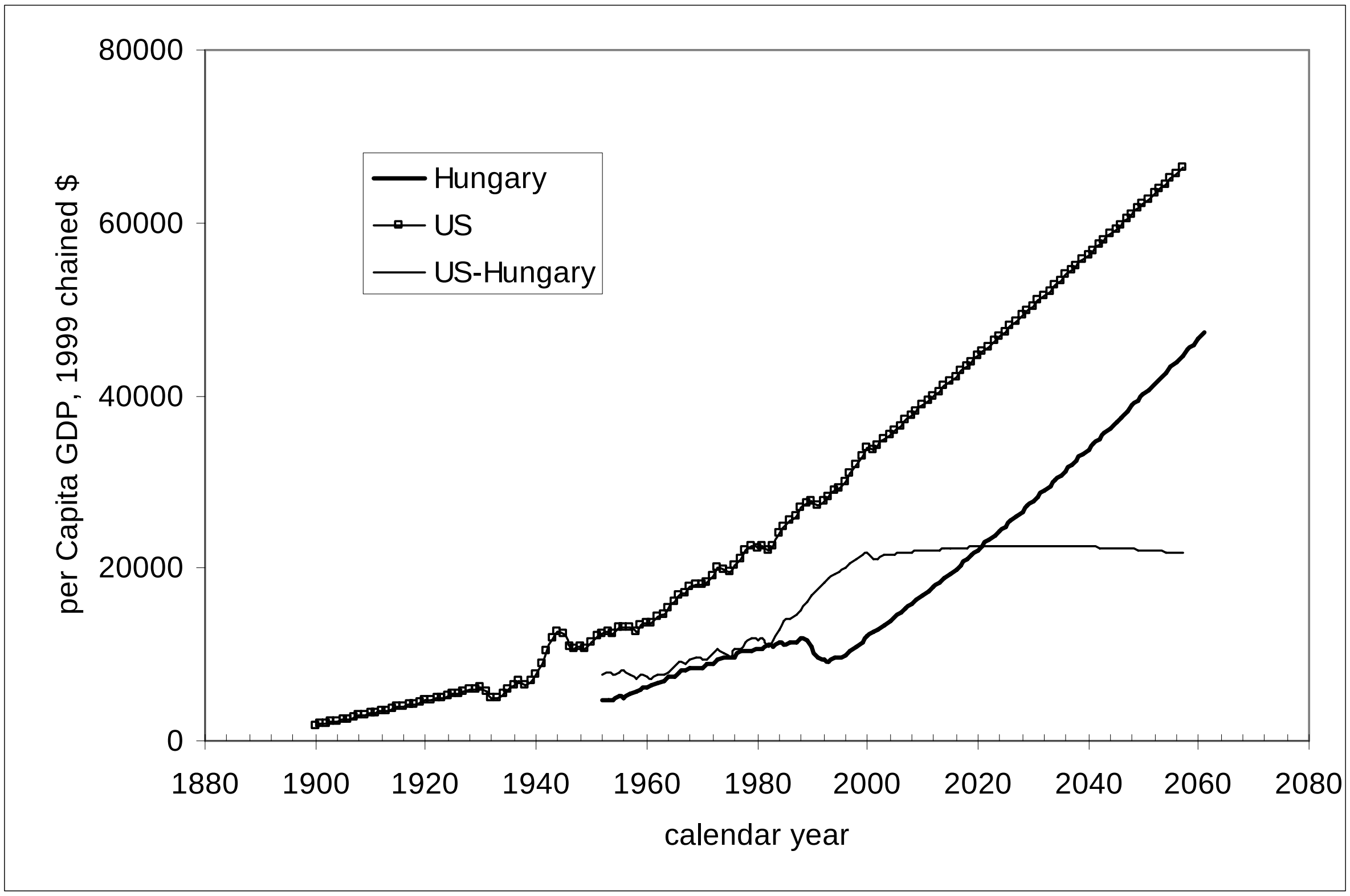

Fig. 15. Future evolution of the per capita GDP in Hungary in comparison with that in the USA. The difference between the countries is also shown (US-Hungary). The best case scenario is that Hungary performs at the theoretical level. The absolute lag between the USA and Hungary can theoretically be constant at level $23,000 in future.

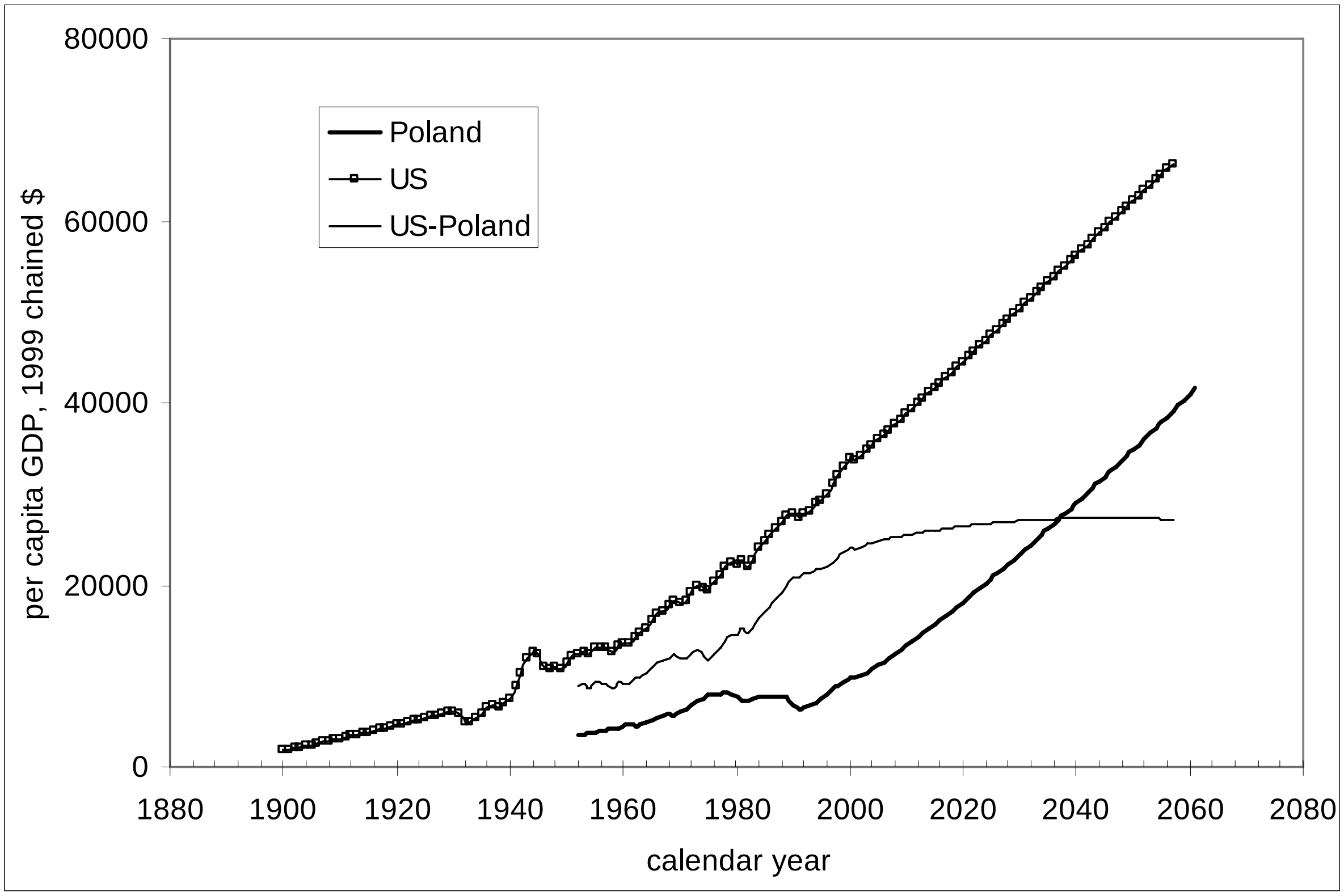

Fig. 16. Future evolution of the per capita GDP in Poland in comparison with that in the USA. The difference between the countries is also shown (US-Poland). The best case scenario is that Poland performs at the theoretical level. The absolute lag between the USA and Poland can theoretically stabilize at about $25,000 (1999) in future.

## Appendix A.

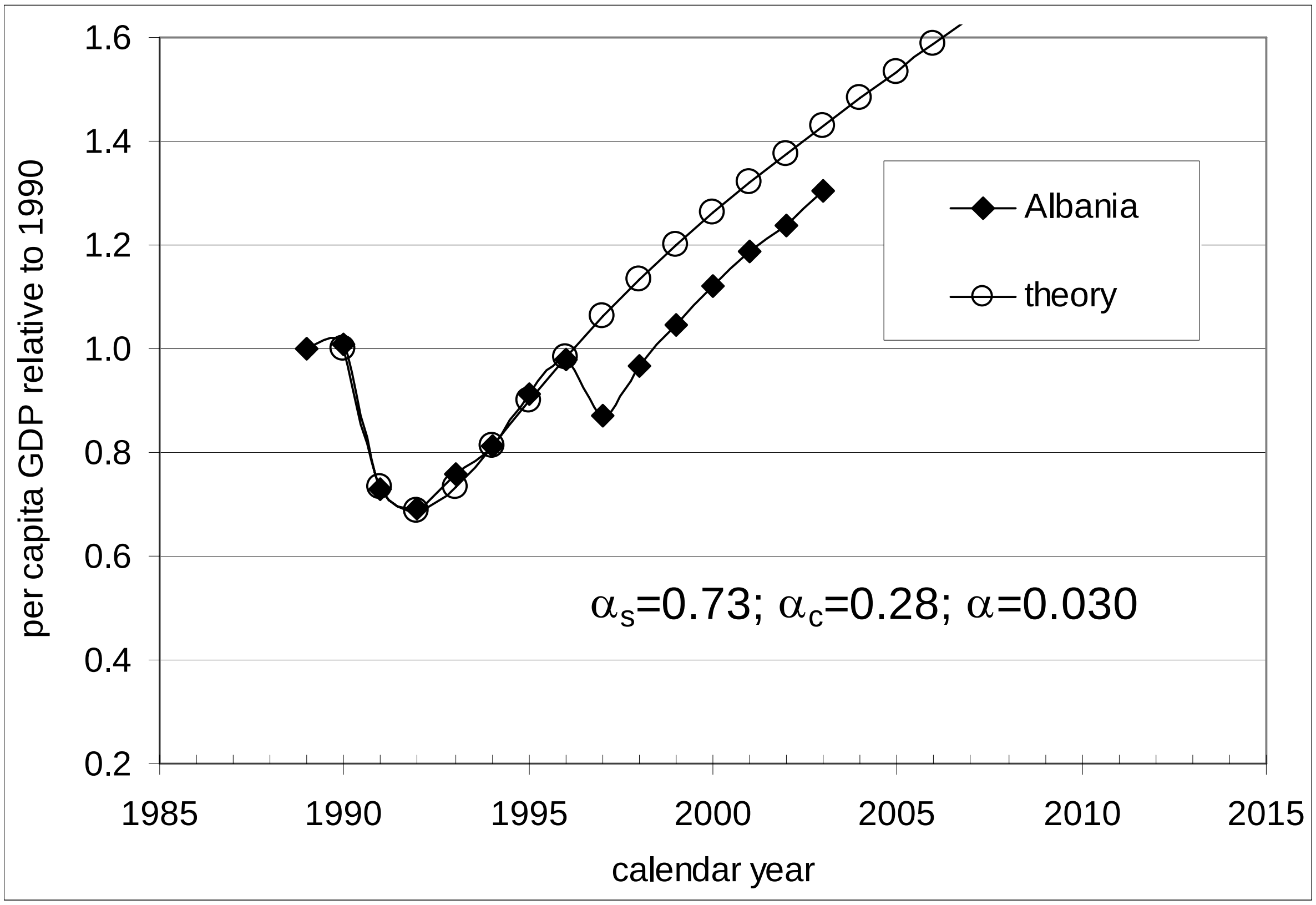

Fig. A1. Comparison of the observed and predicted transition process for the replacement of the socialist system with the capitalist system in Albania. Parameters of the model are indicated.

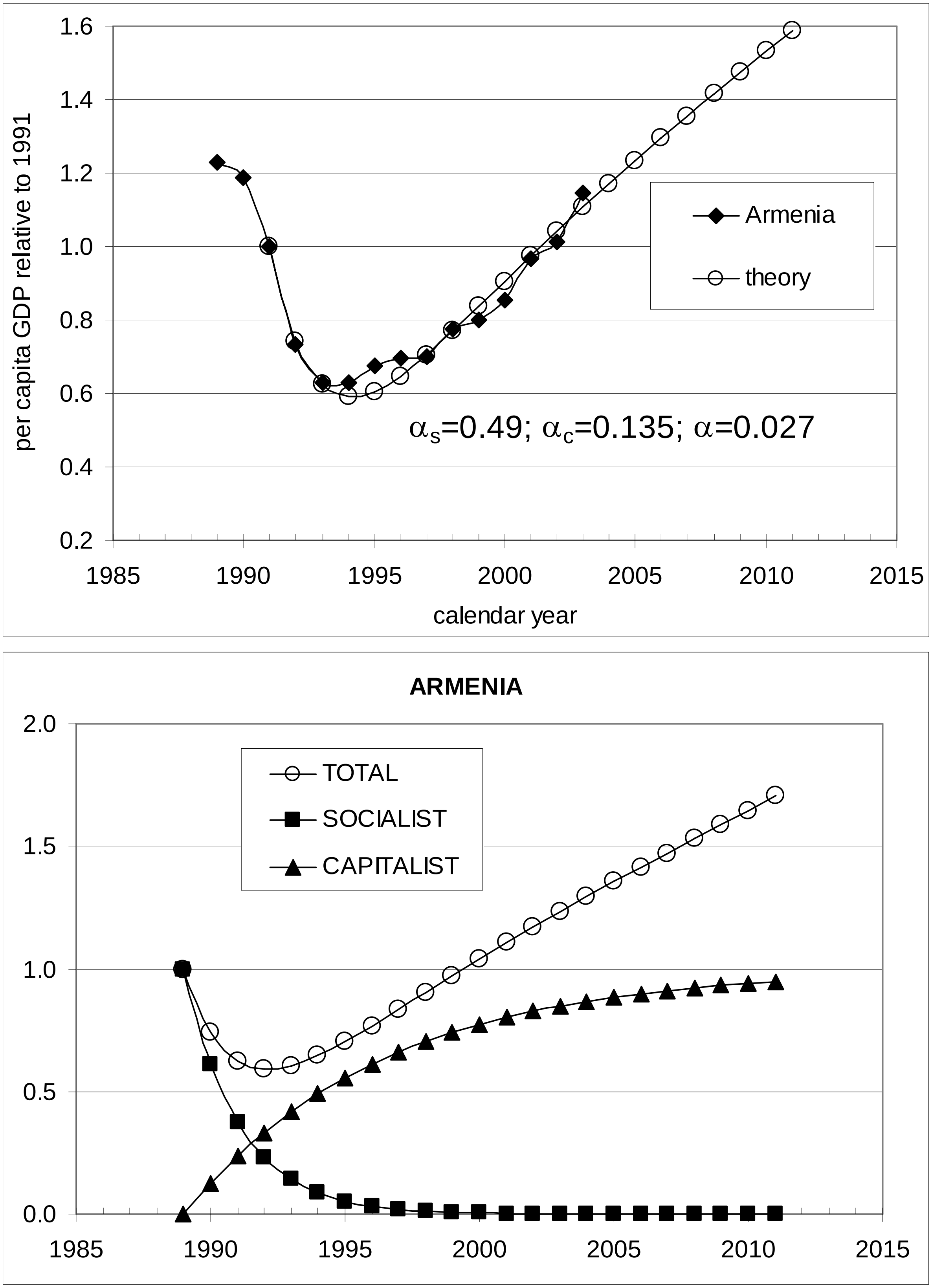

Fig. A2. Comparison of the observed and predicted transition process for the replacement of the socialist system with the capitalist system in Armenia. Parameters of the model are indicated.

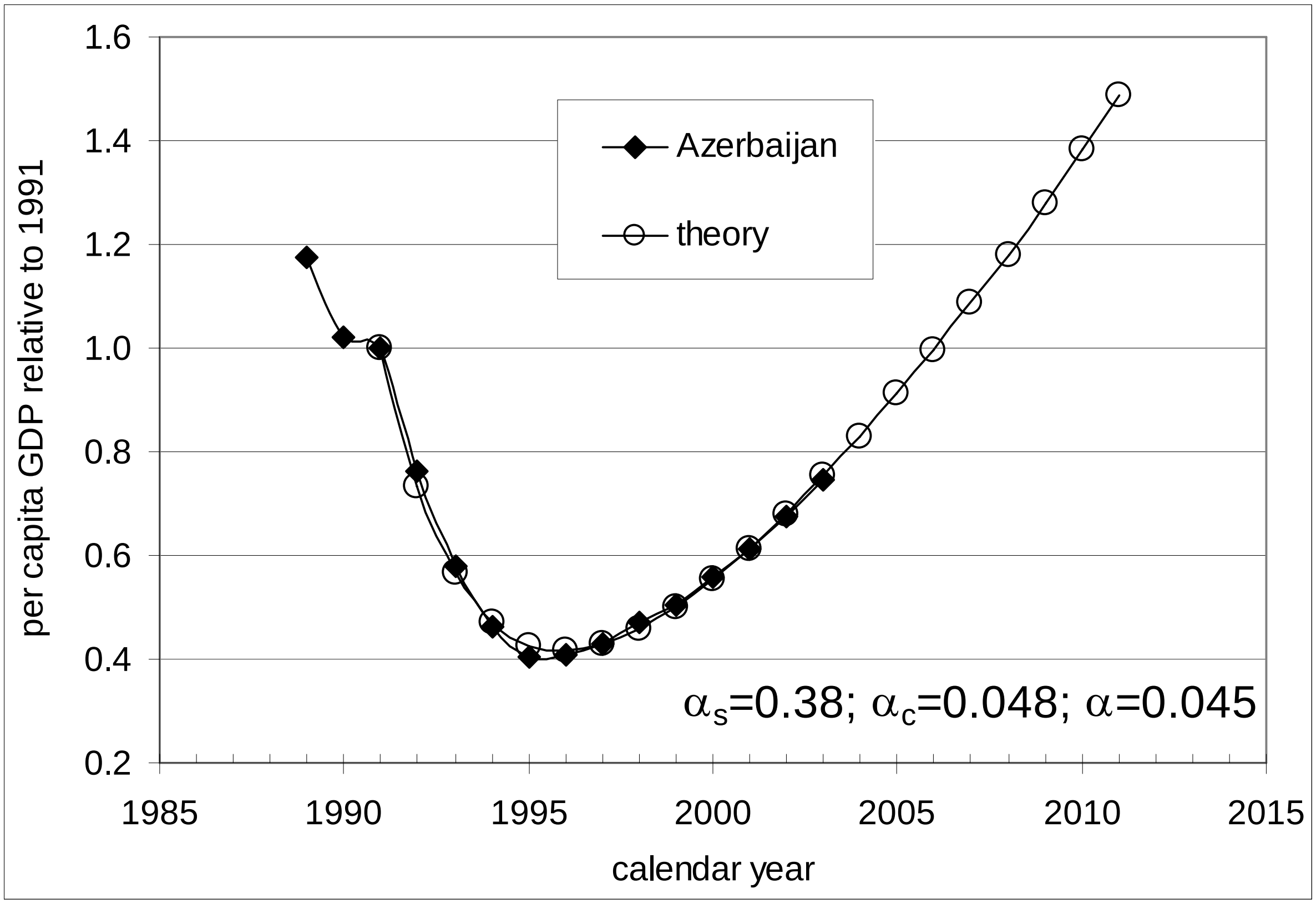

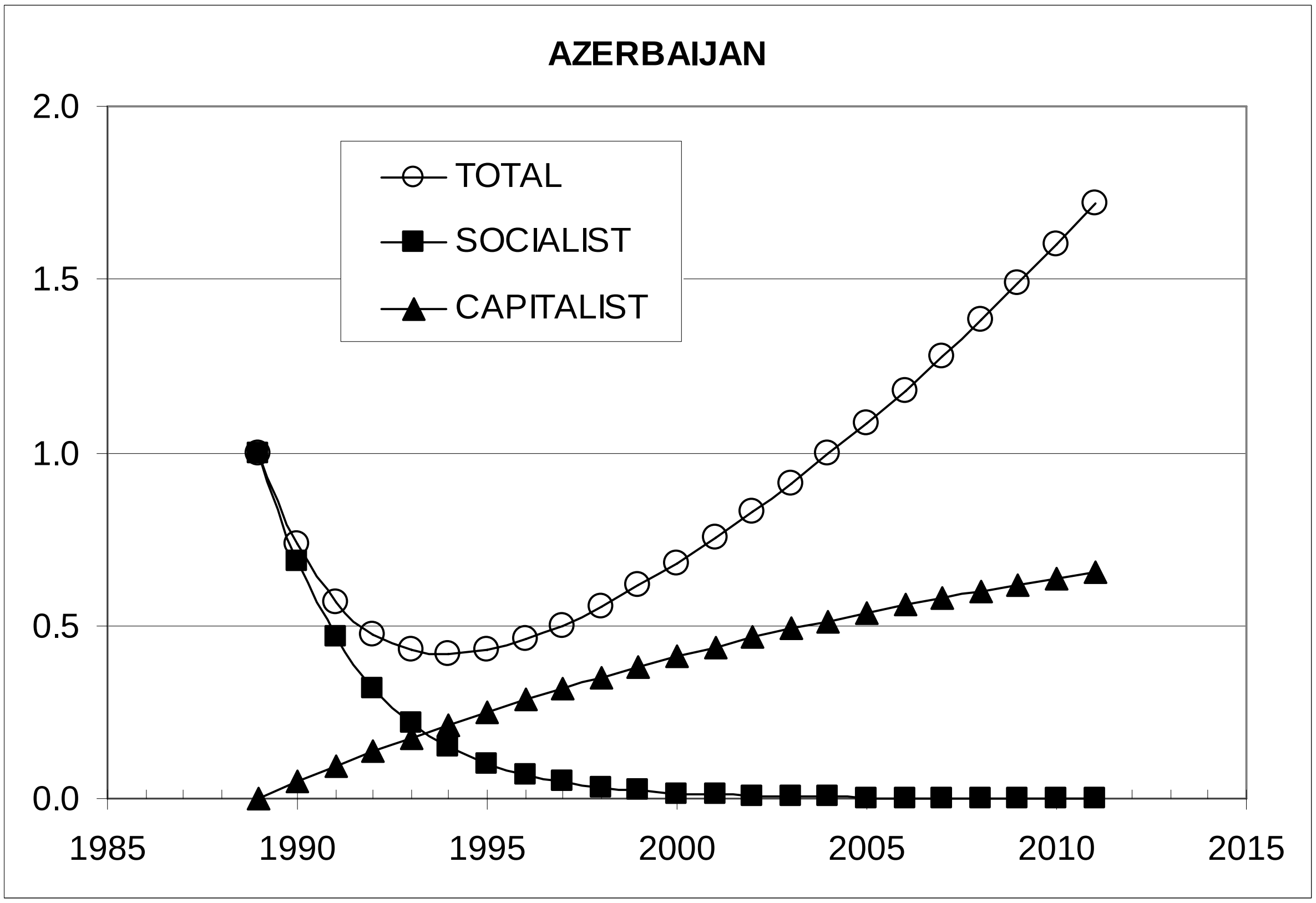

Fig. A3. Comparison of the observed and predicted transition process for the replacement of the socialist system with the capitalist system in Azerbaijan. Parameters of the model are indicated.

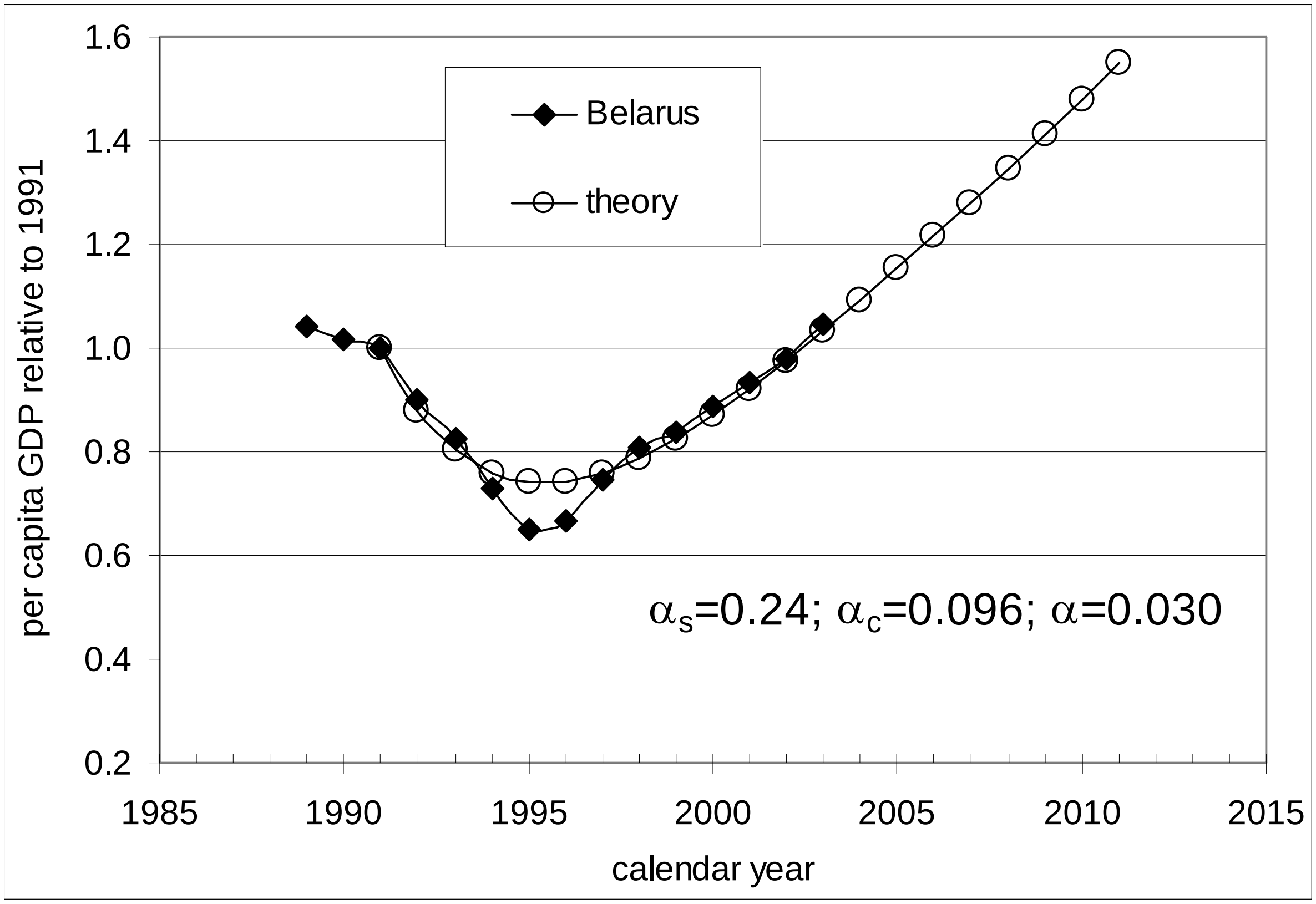

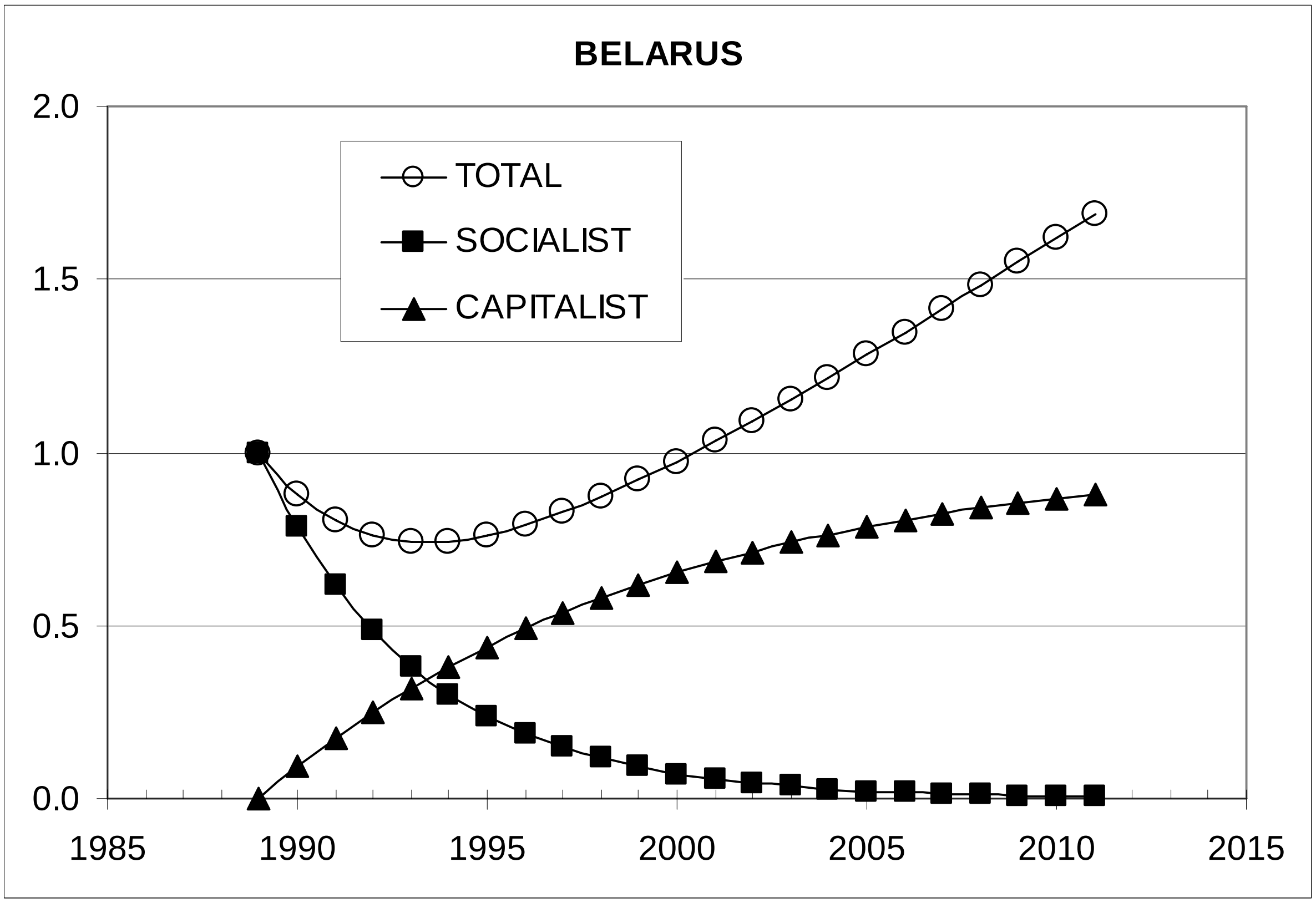

Fig. A4. Comparison of the observed and predicted transition process for the replacement of the socialist system with the capitalist system in Belarus. Parameters of the model are indicated.

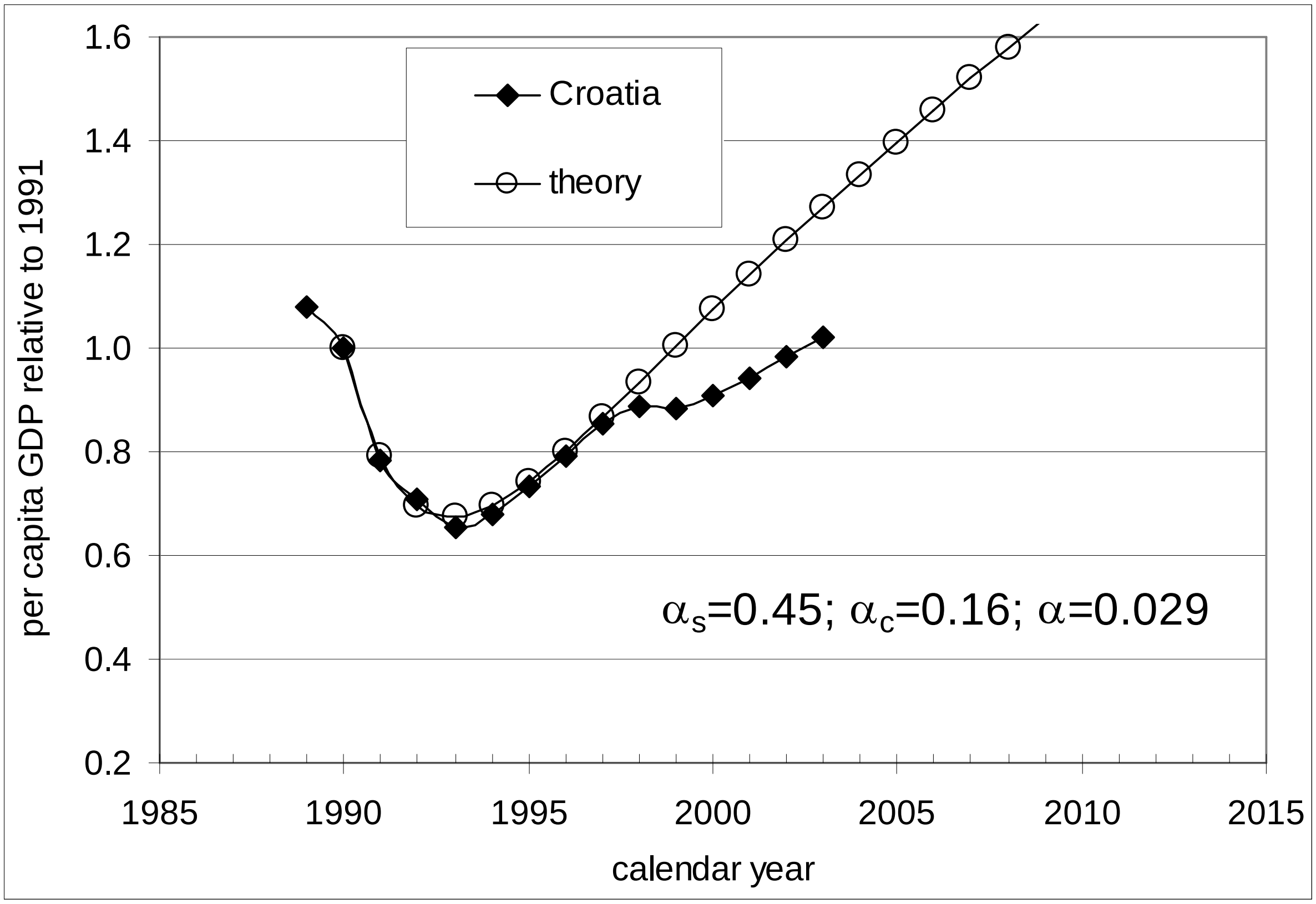

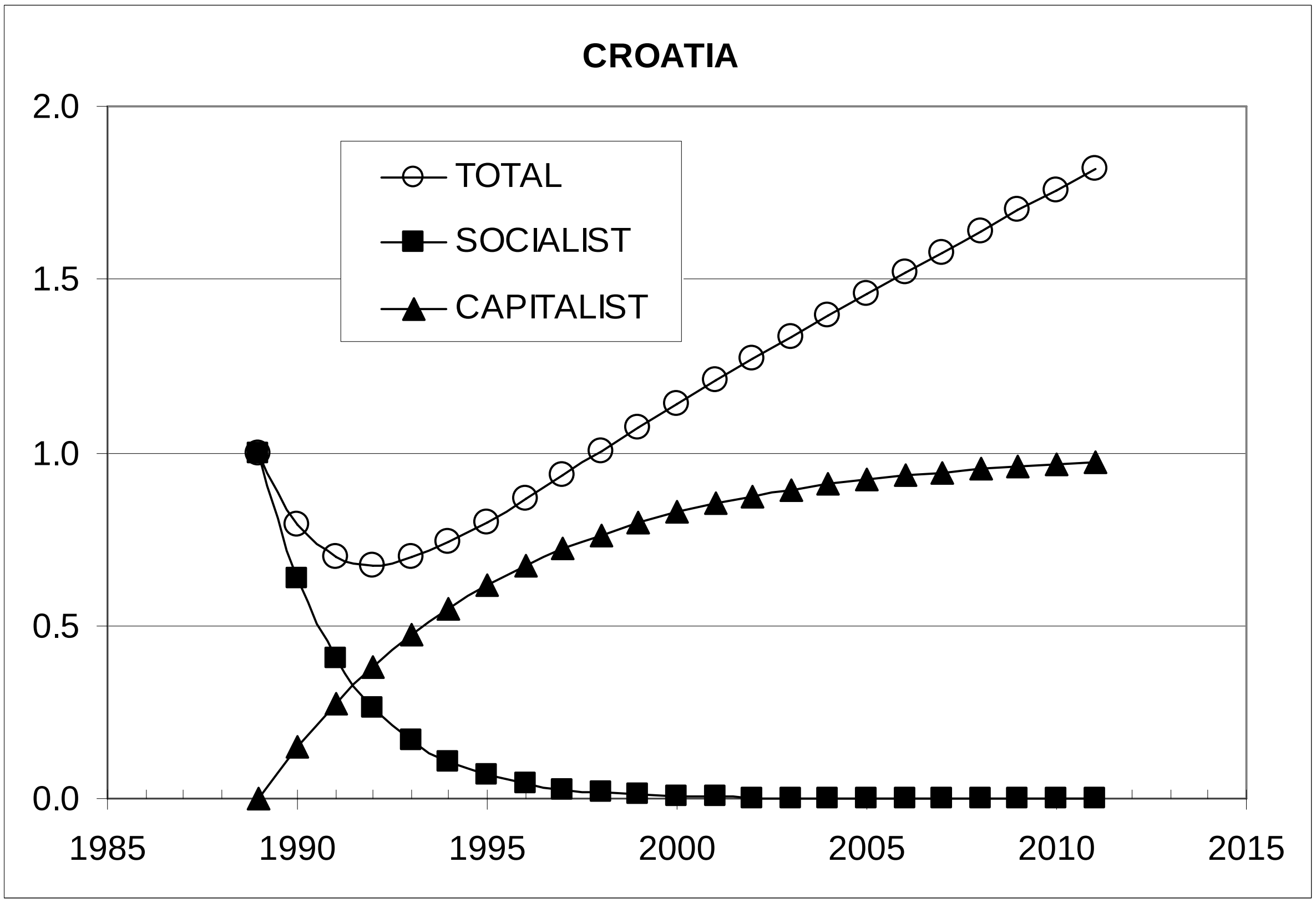

Fig. A5. Comparison of the observed and predicted transition process for the replacement of the socialist system with the capitalist system in Croatia. Parameters of the model are indicated.

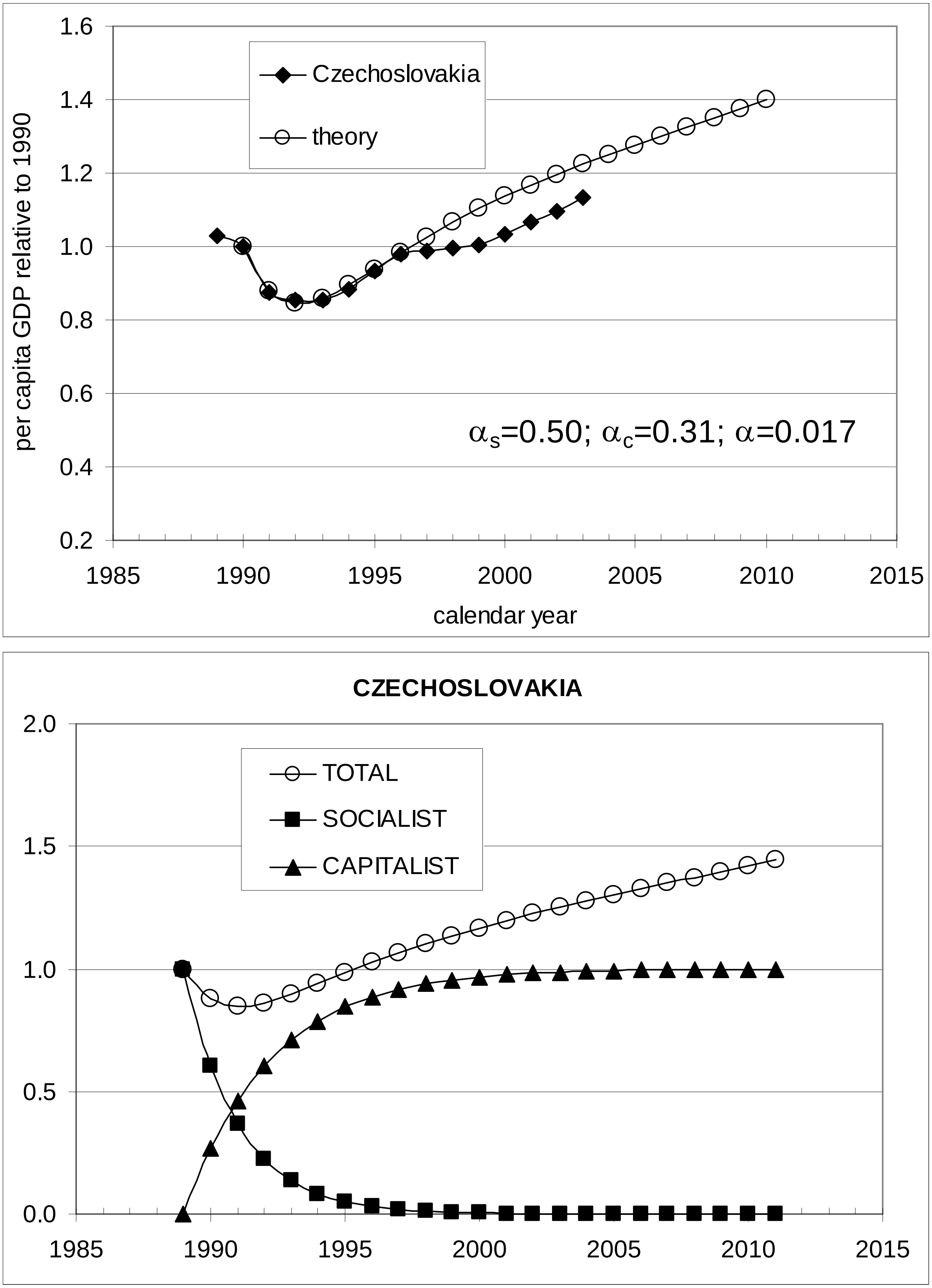

Fig. A6. Comparison of the observed and predicted transition process for the replacement of the socialist system with the capitalist system in Czechoslovakia. Parameters of the model are indicated.

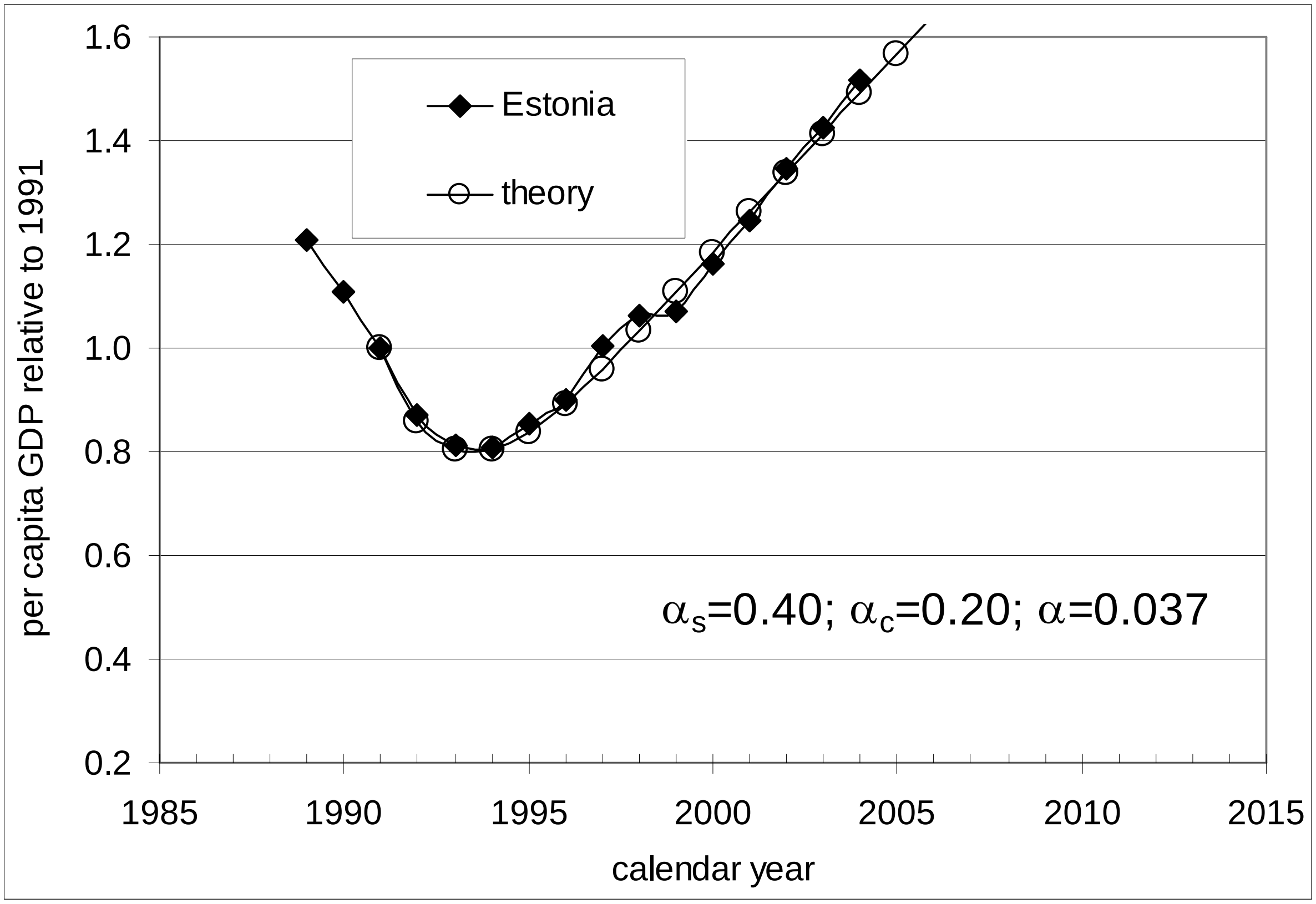

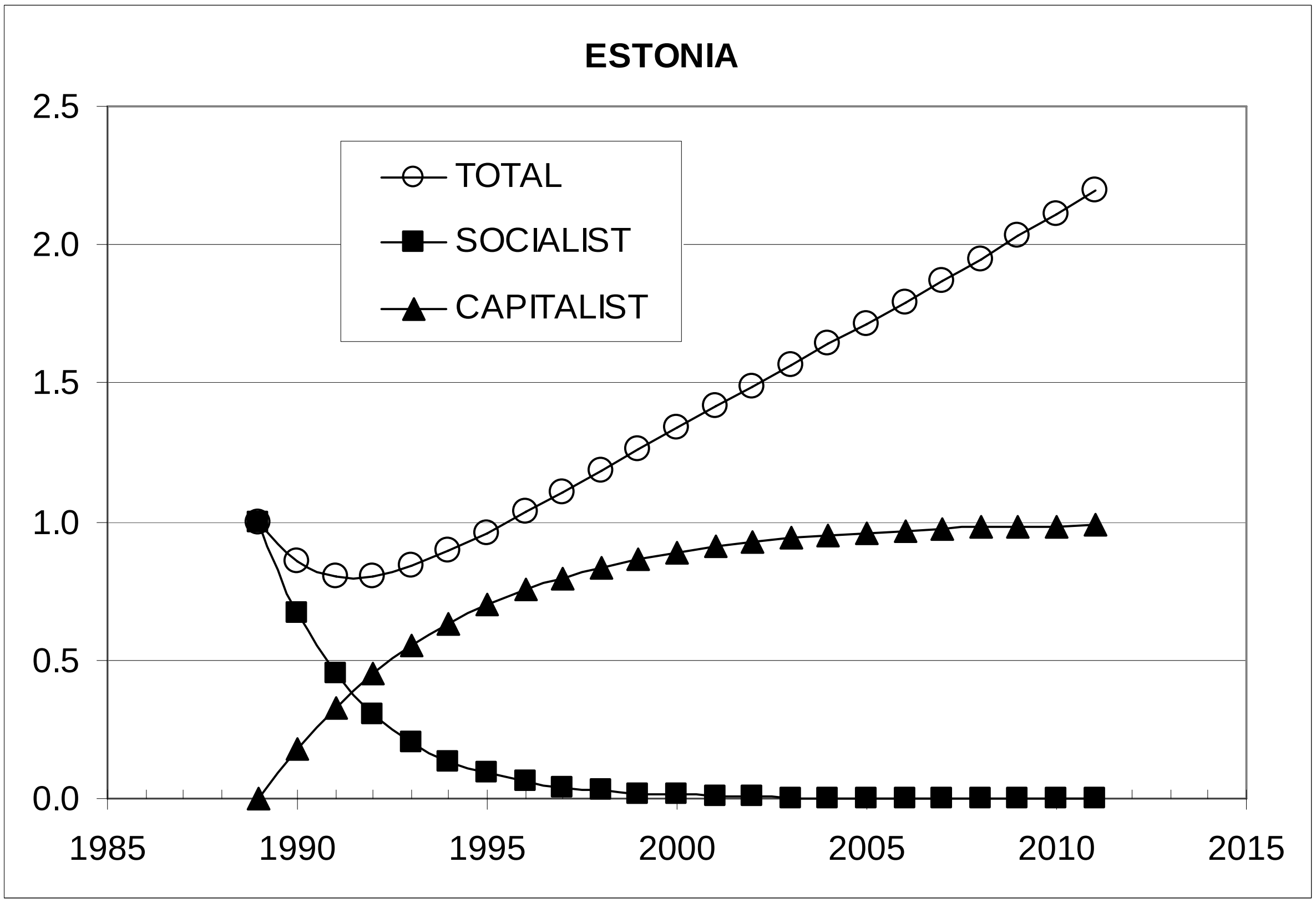

Fig. A7. Comparison of the observed and predicted transition process for the replacement of the socialist system with the capitalist system in Estonia. Parameters of the model are indicated.

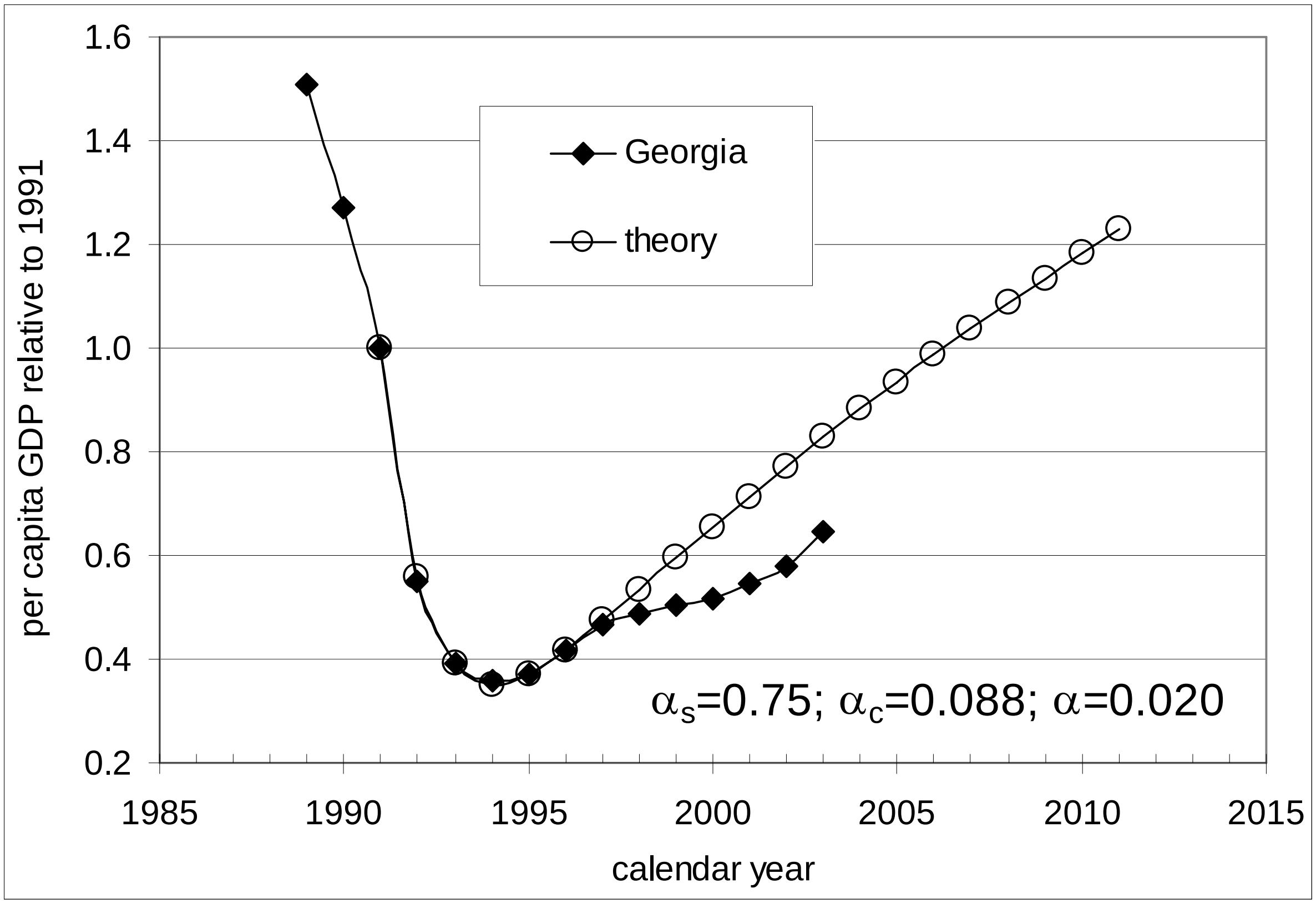

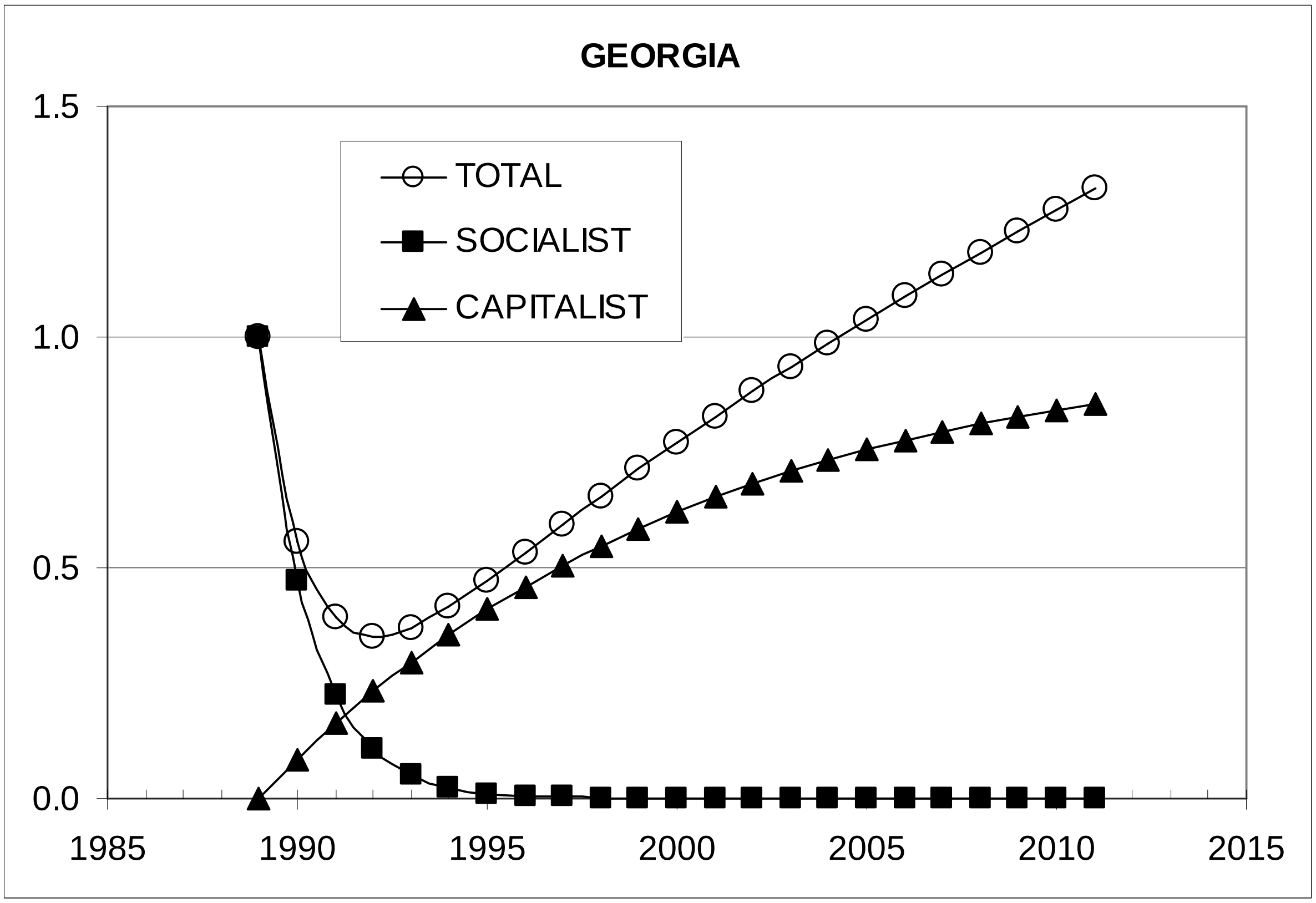

Fig. A8. Comparison of the observed and predicted transition process for the replacement of the socialist system with the capitalist system in Georgia. Parameters of the model are indicated.

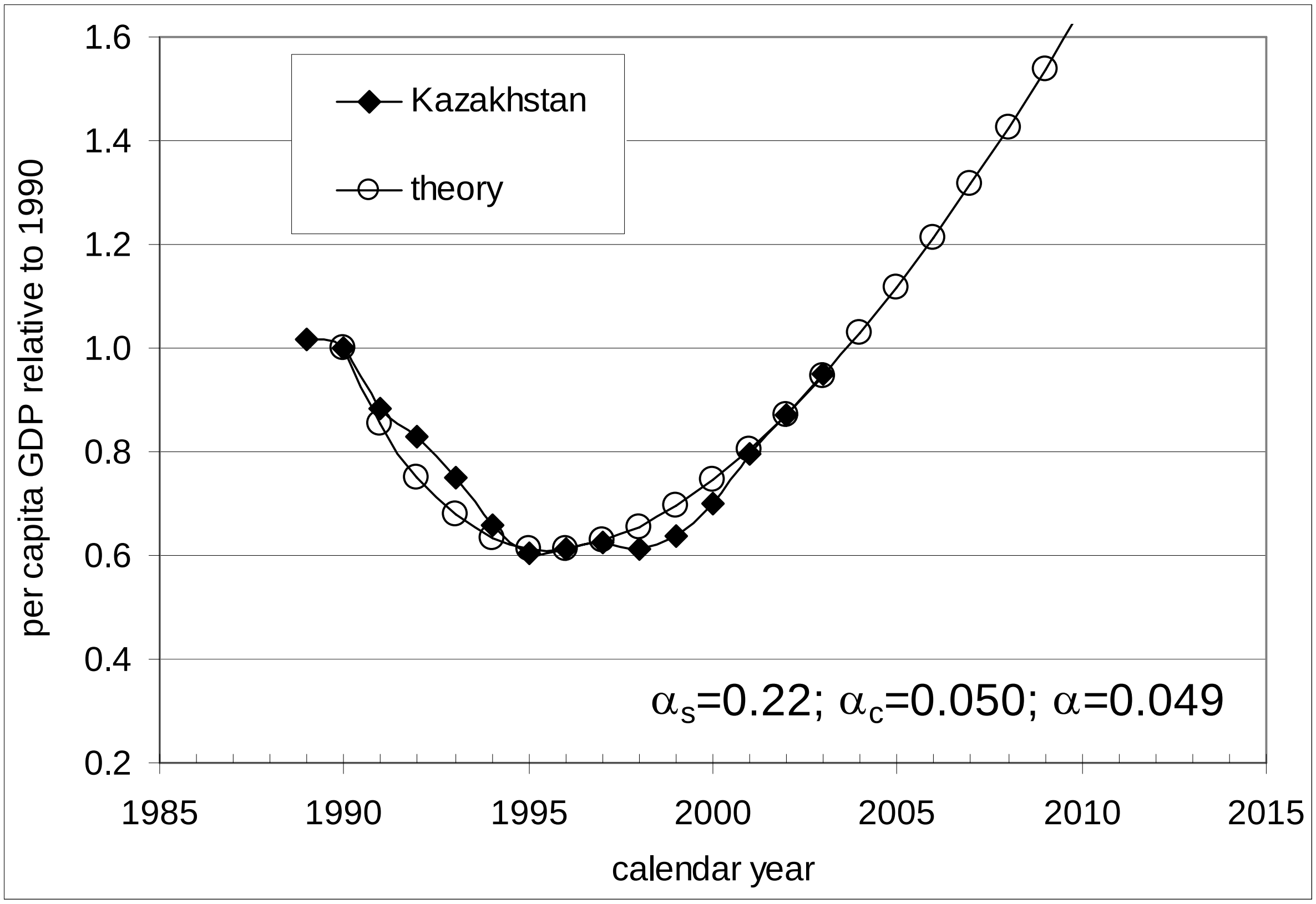

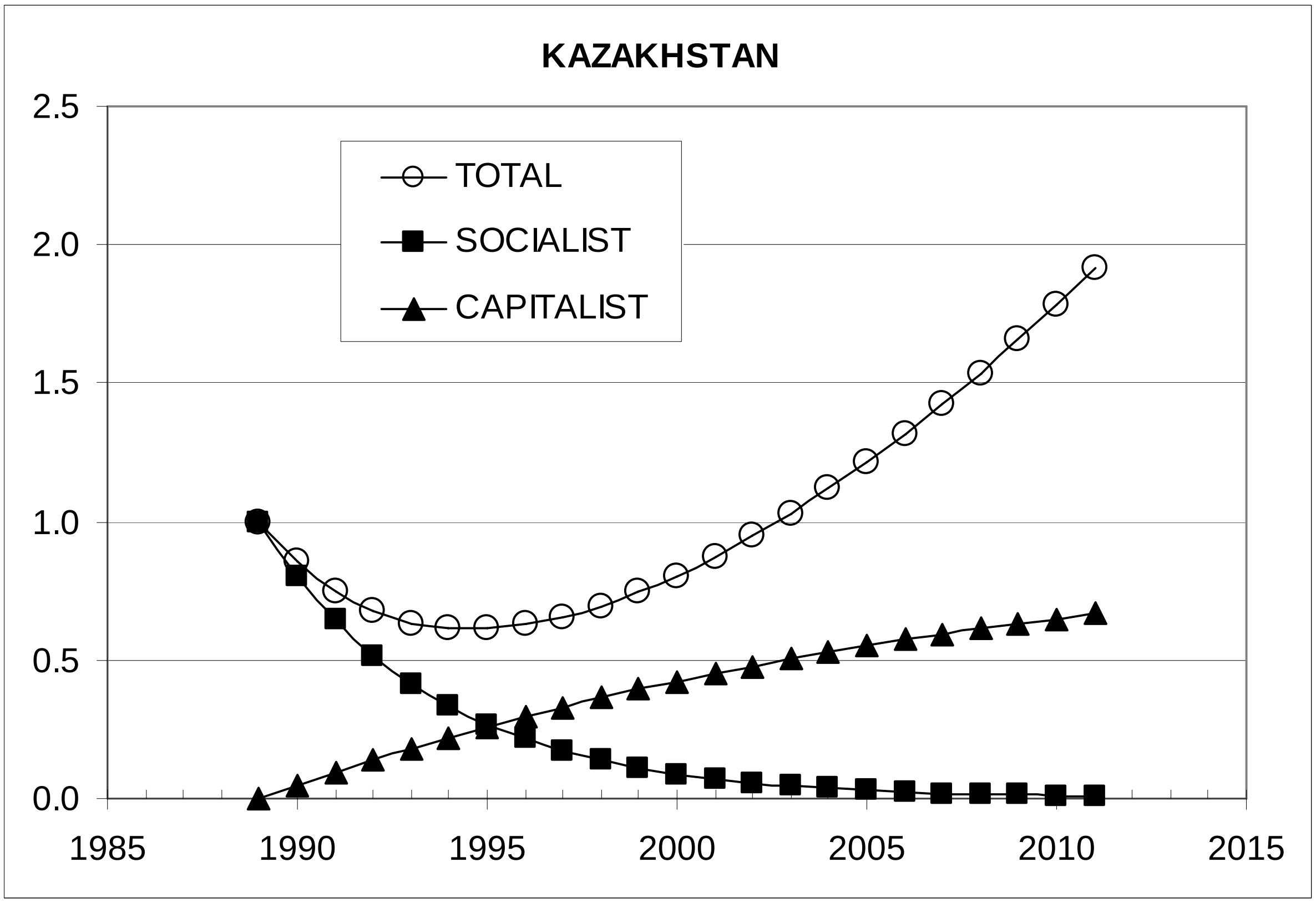

Fig. A9. Comparison of the observed and predicted transition process for the replacement of the socialist system with the capitalist system in Kazakhstan. Parameters of the model are indicated.

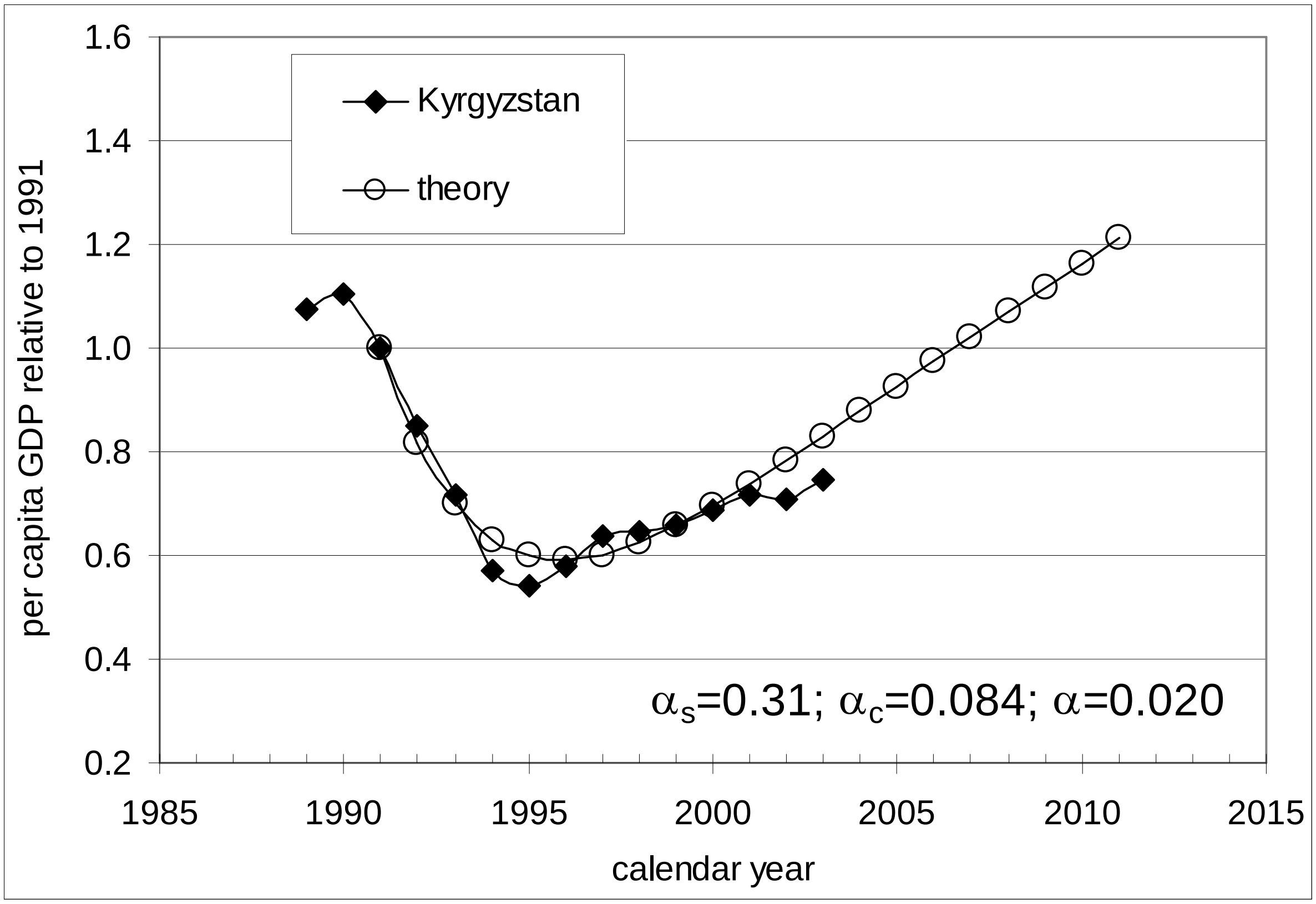

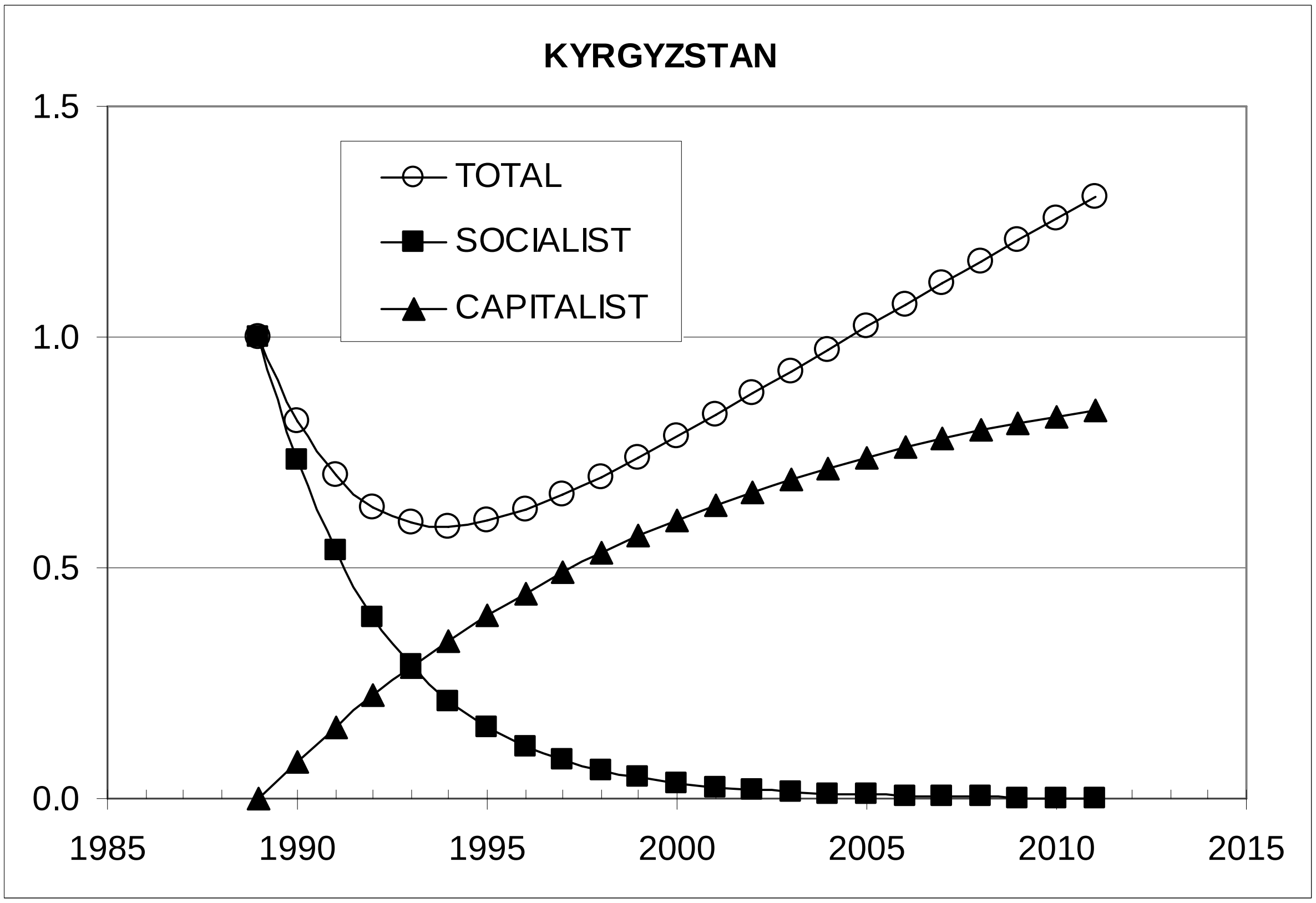

Fig. A10. Comparison of the observed and predicted transition process for the replacement of the socialist system with the capitalist system in Kyrgyz. Parameters of the model are indicated.

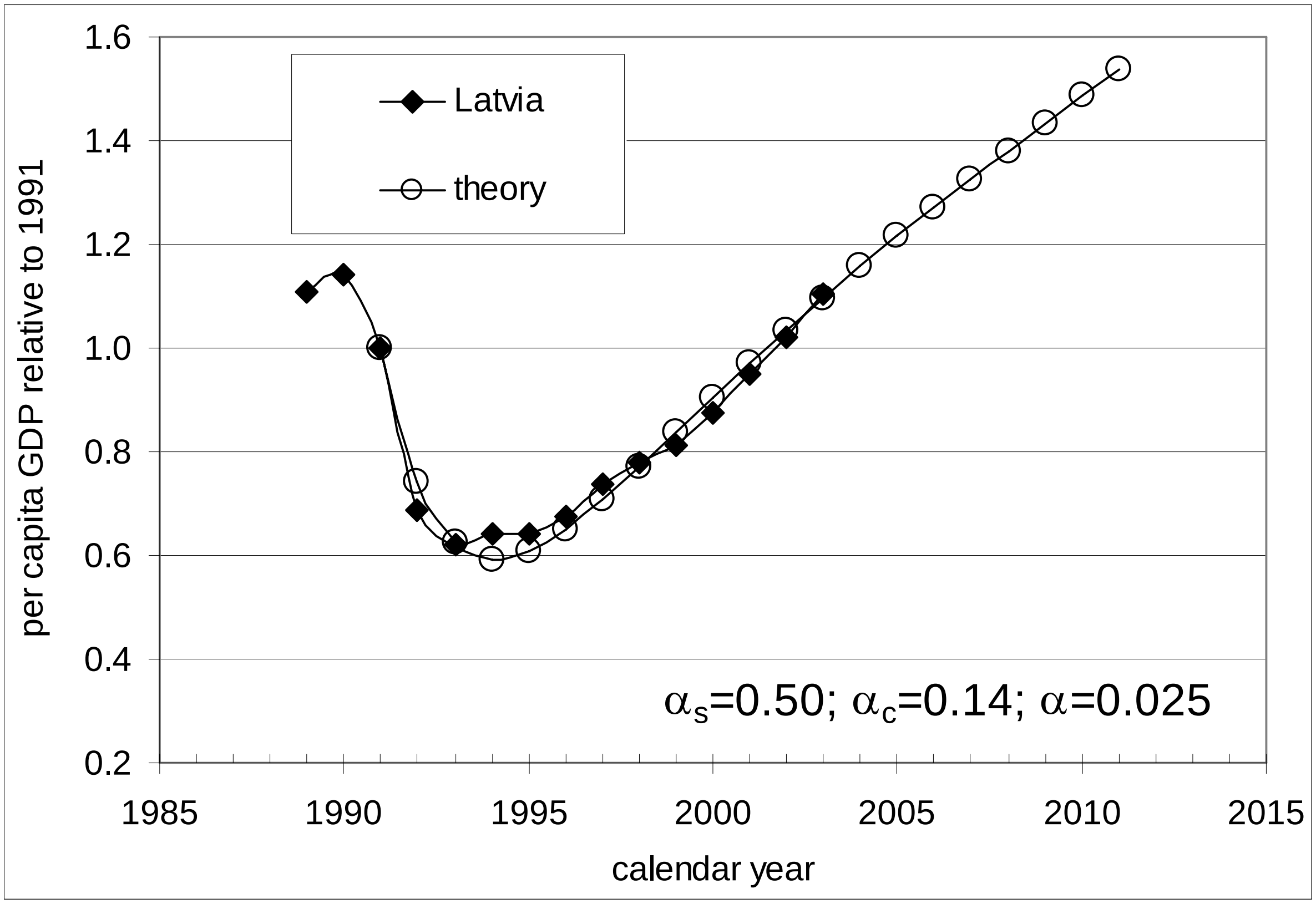

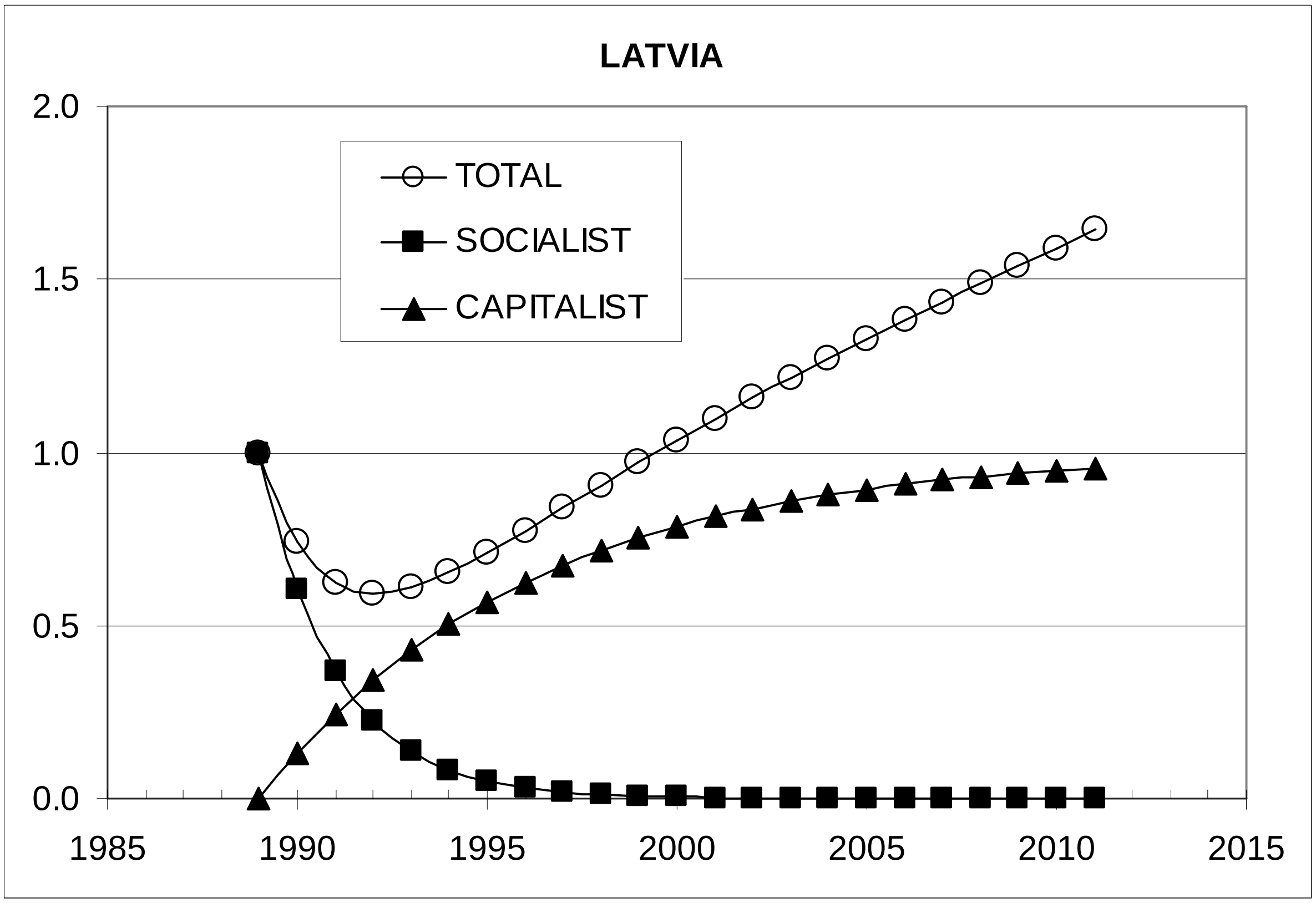

Fig. A11. Comparison of the observed and predicted transition process for the replacement of the socialist system with the capitalist system in Latvia. Parameters of the model are indicated.

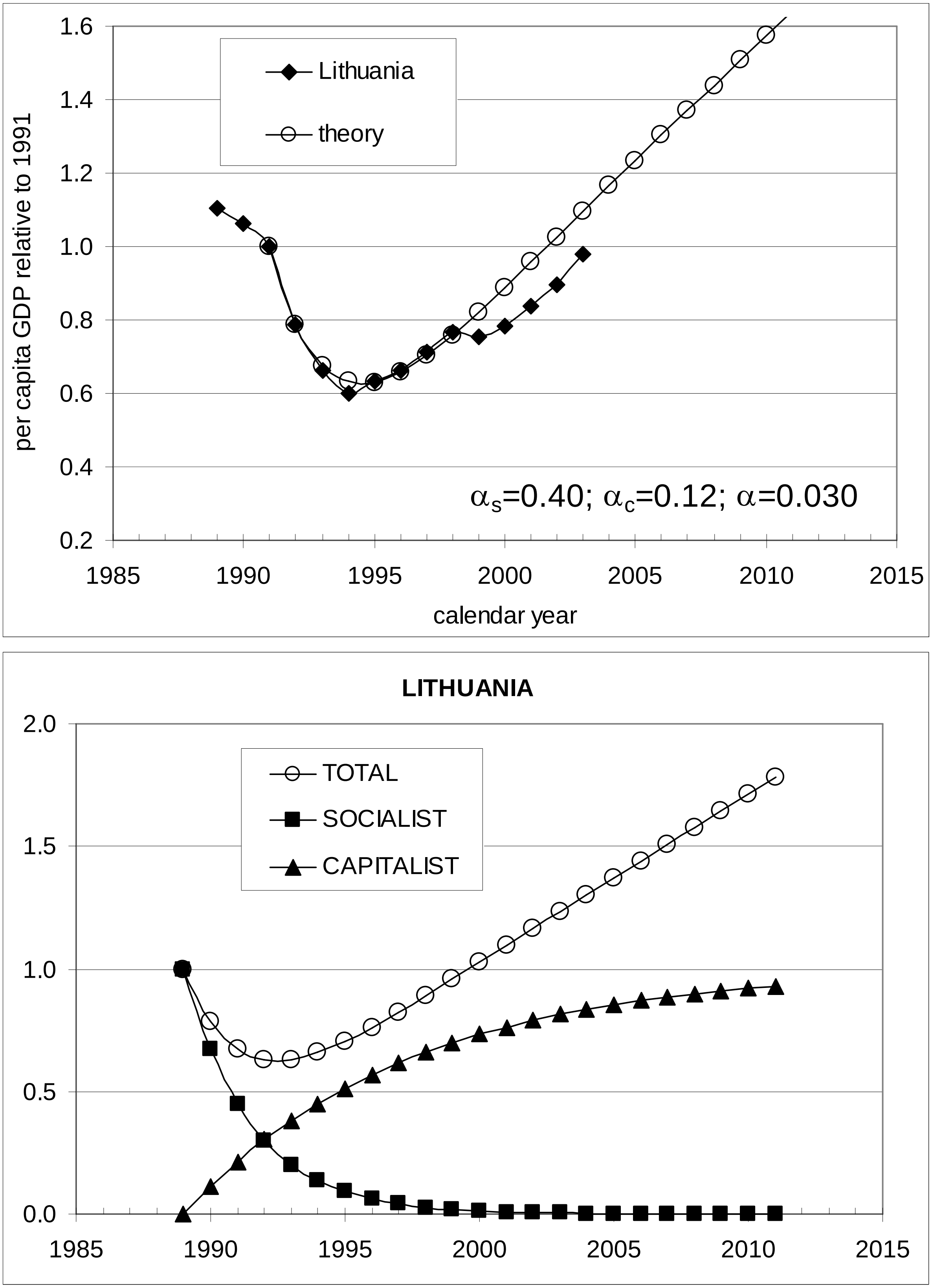

Fig. A12. Comparison of the observed and predicted transition process for the replacement of the socialist system with the capitalist system in Lithuania. Parameters of the model are indicated.

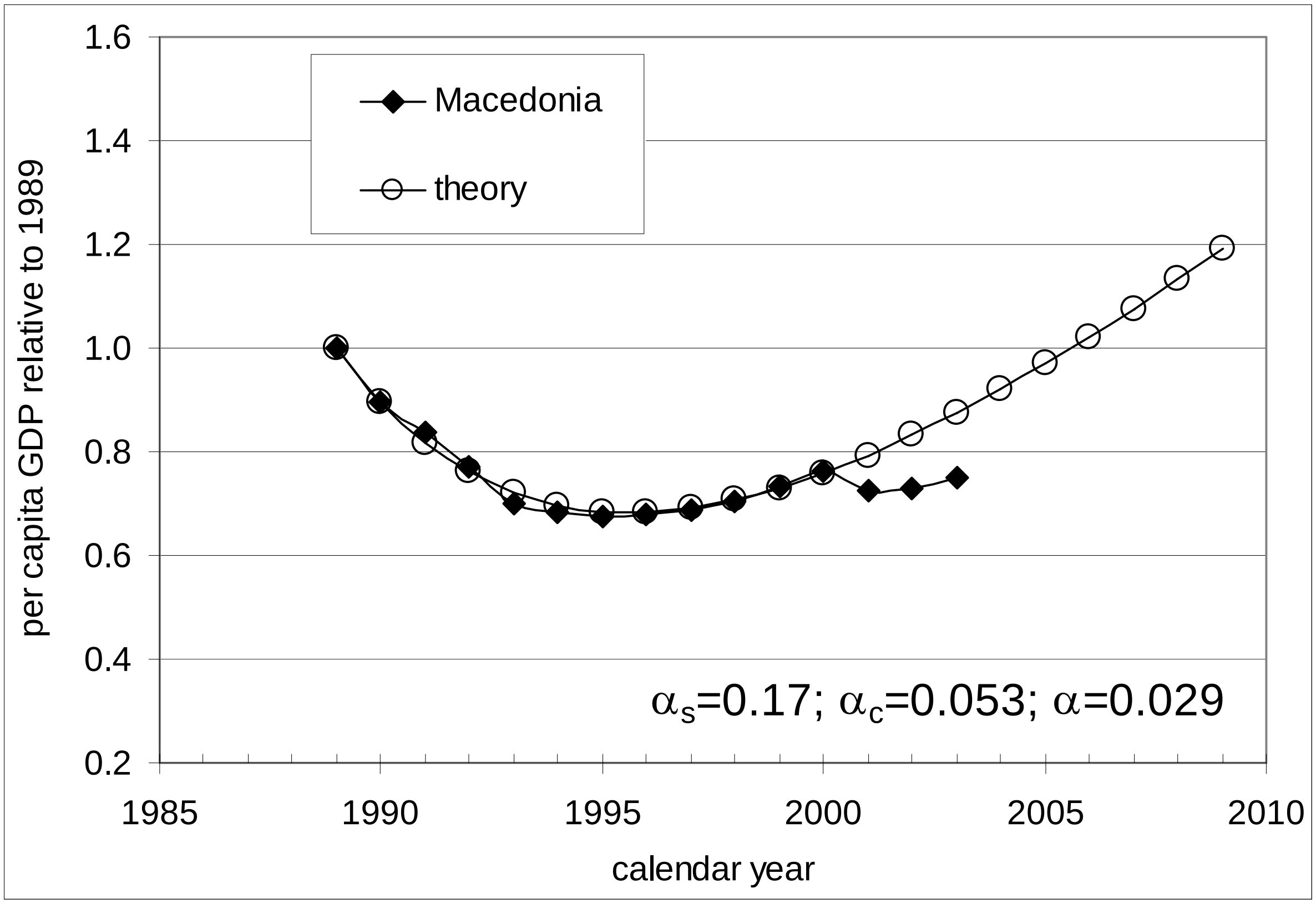

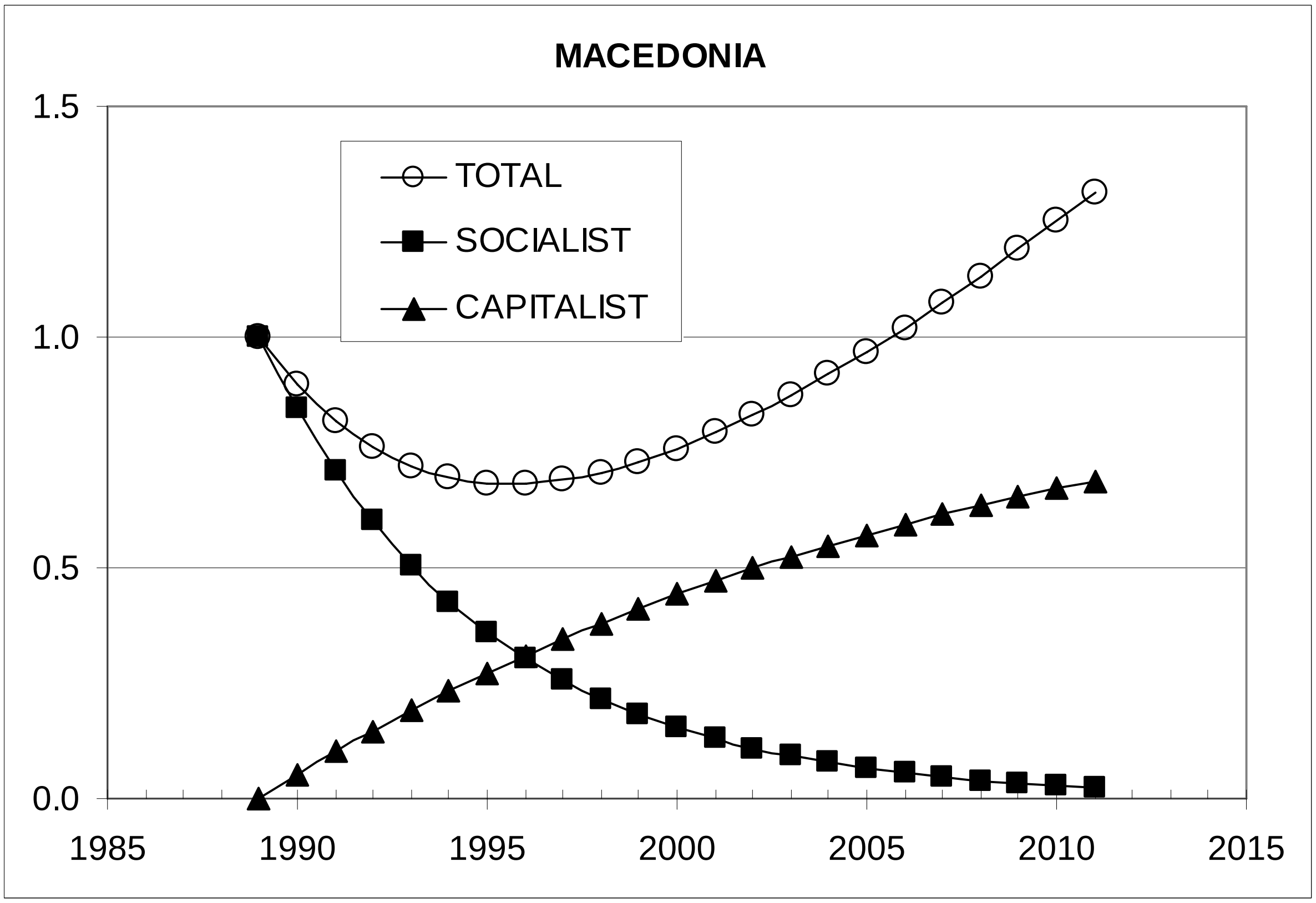

Fig. A13. Comparison of the observed and predicted transition process for the replacement of the socialist system with the capitalist system in Macedonia. Parameters of the model are indicated.

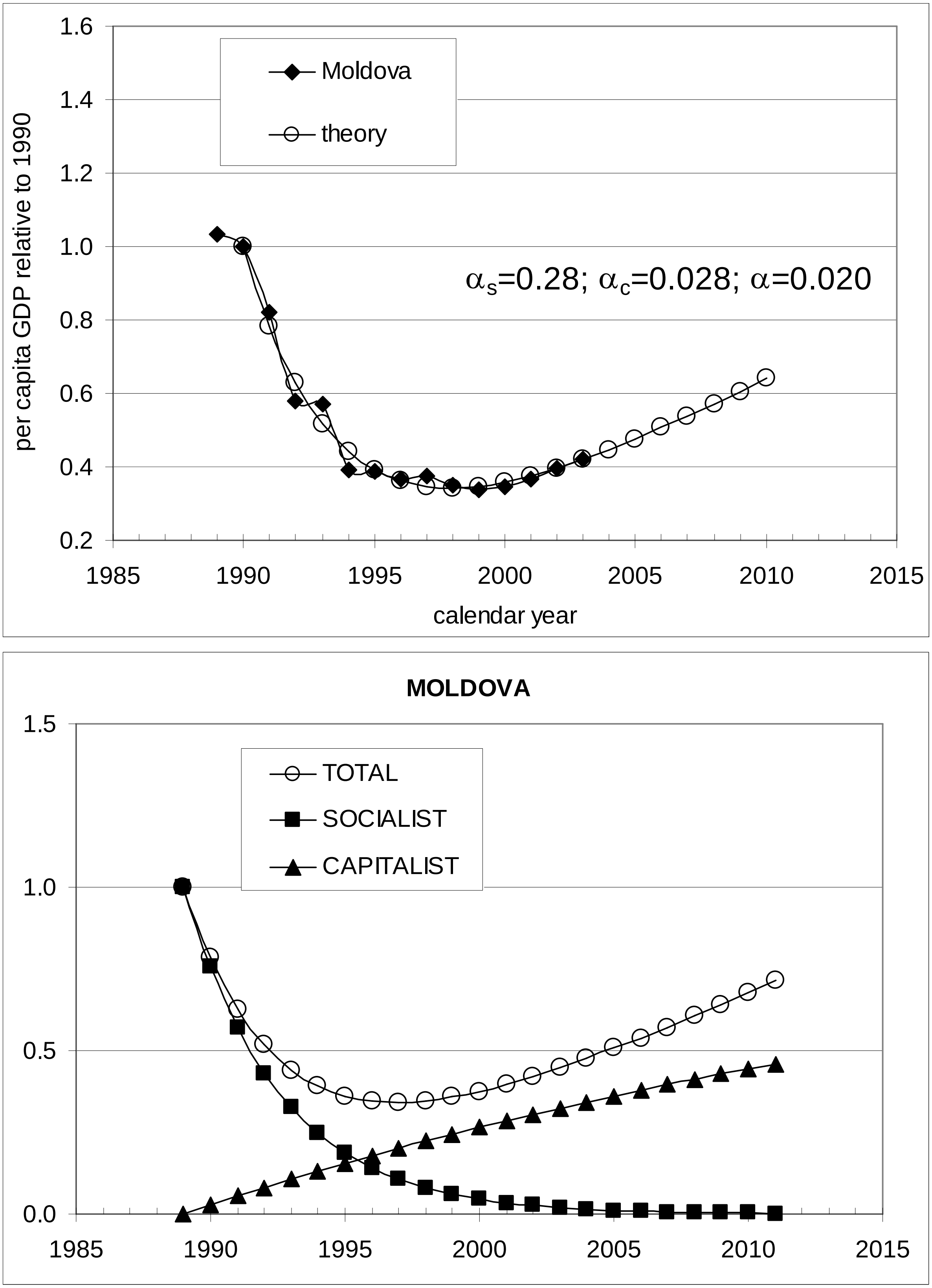

Fig. A14. Comparison of the observed and predicted transition process for the replacement of the socialist system with the capitalist system in Moldova. Parameters of the model are indicated.

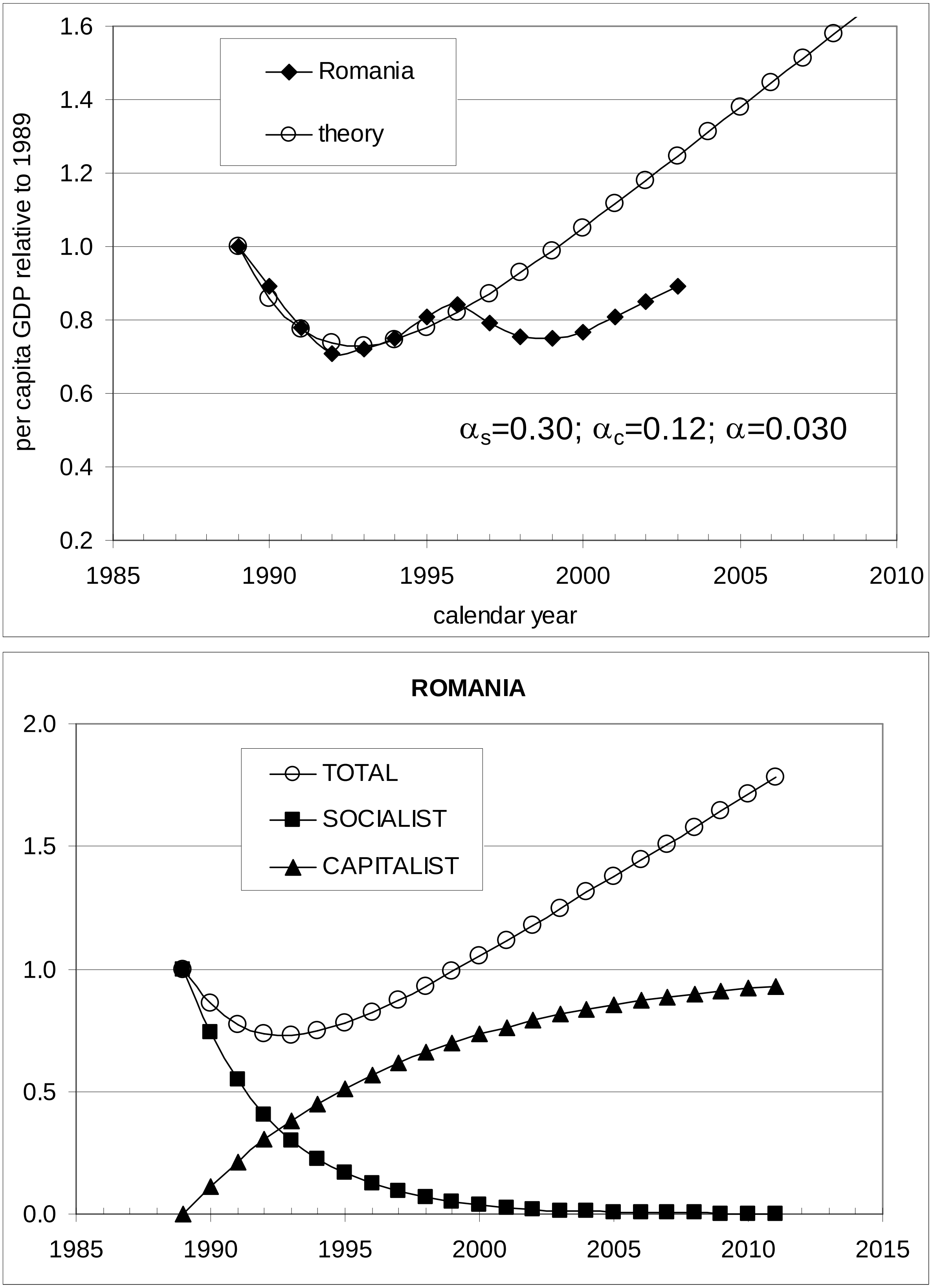

Fig. A15. Comparison of the observed and predicted transition process for the replacement of the socialist system with the capitalist system in Romania. Parameters of the model are indicated.

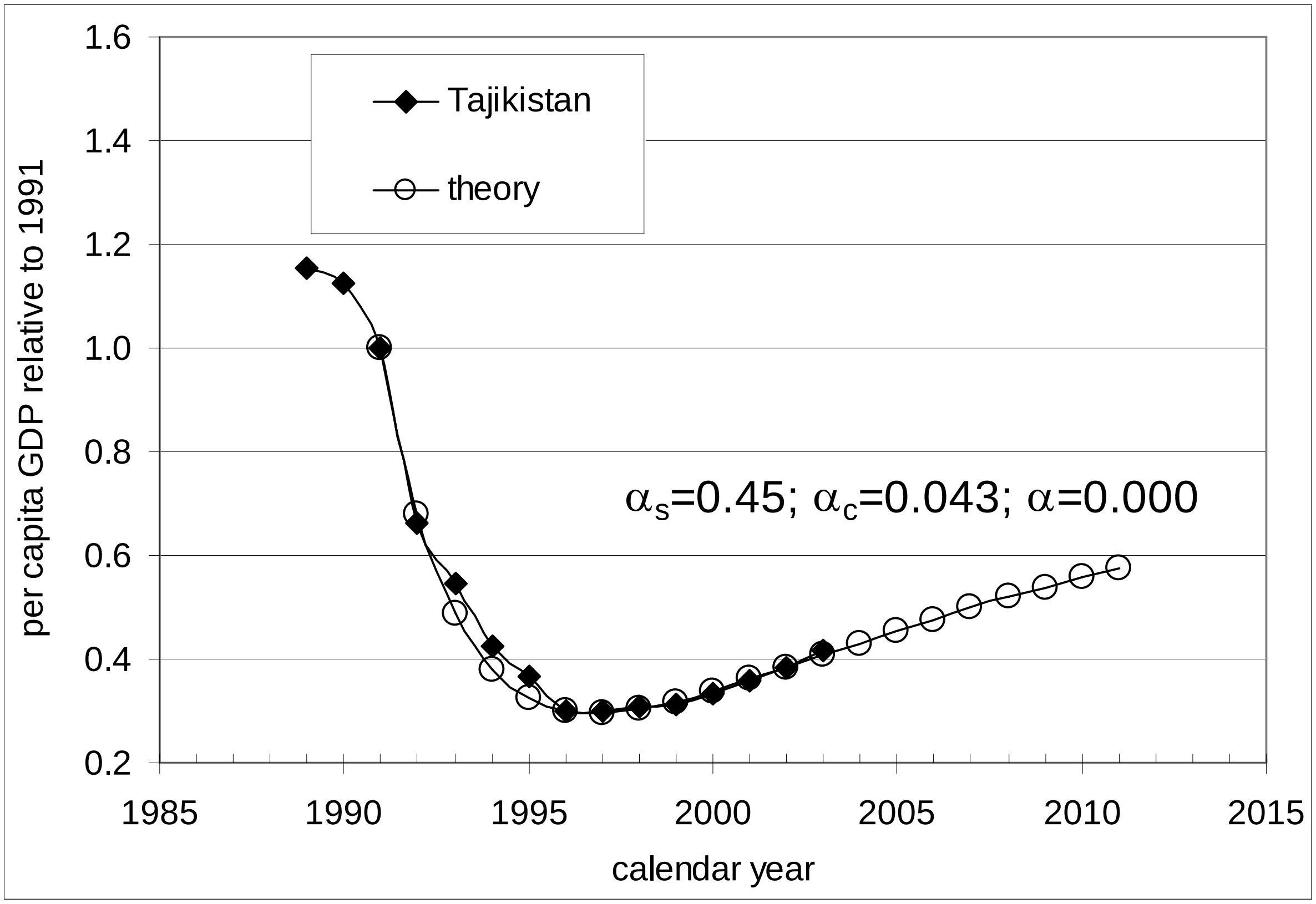

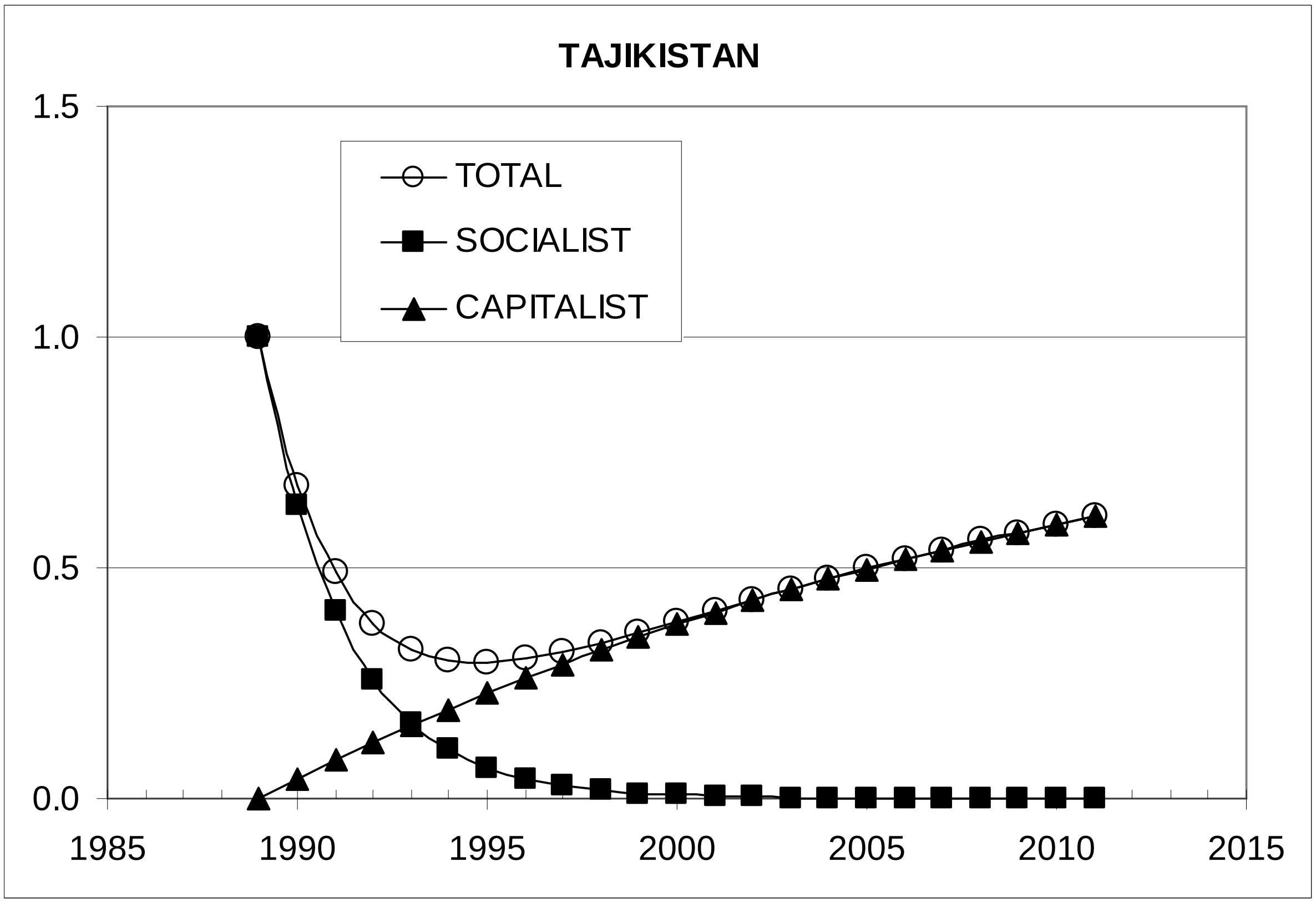

Fig. A16. Comparison of the observed and predicted transition process for the replacement of the socialist system with the capitalist system in Tajikistan. Parameters of the model are indicated.

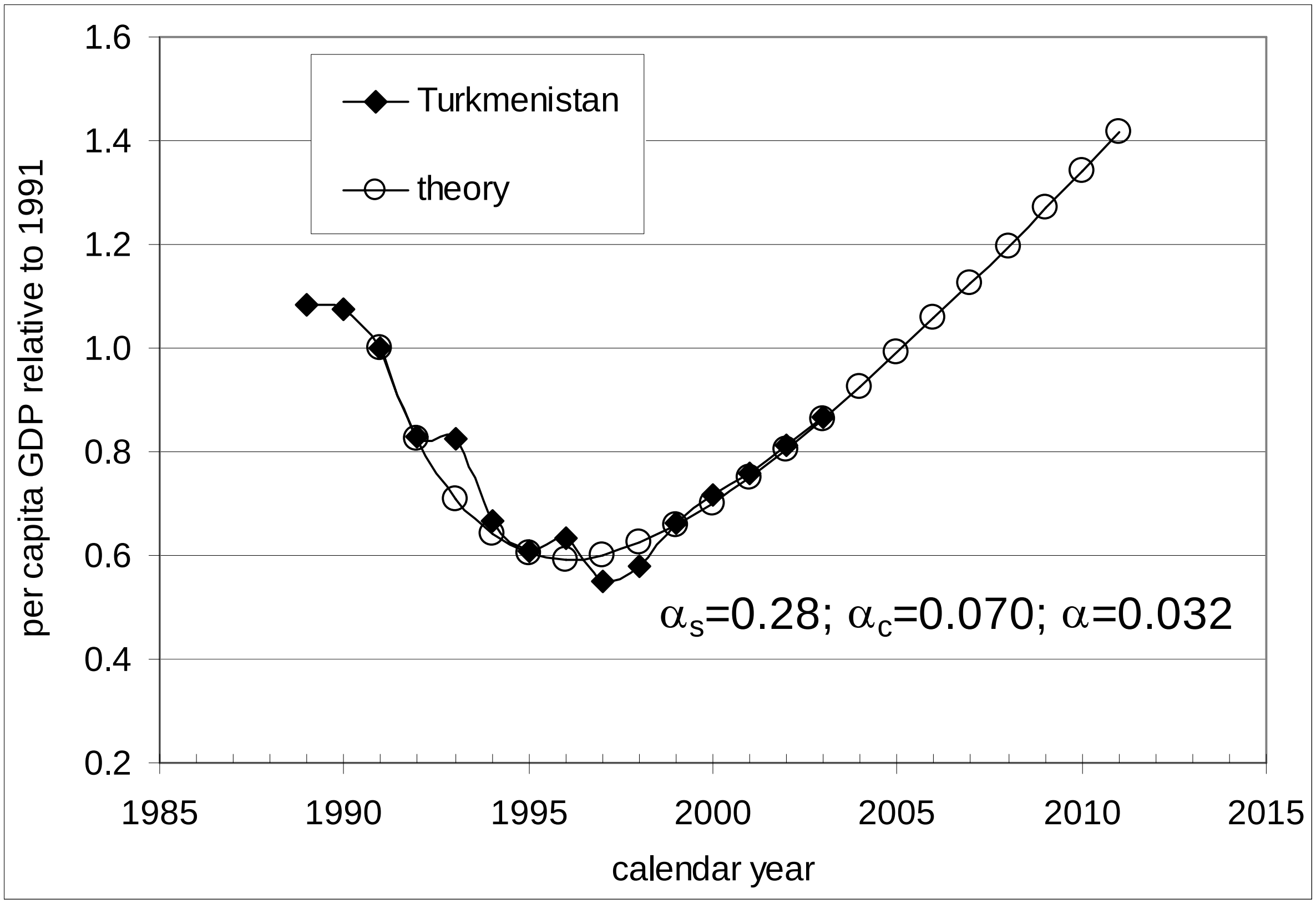

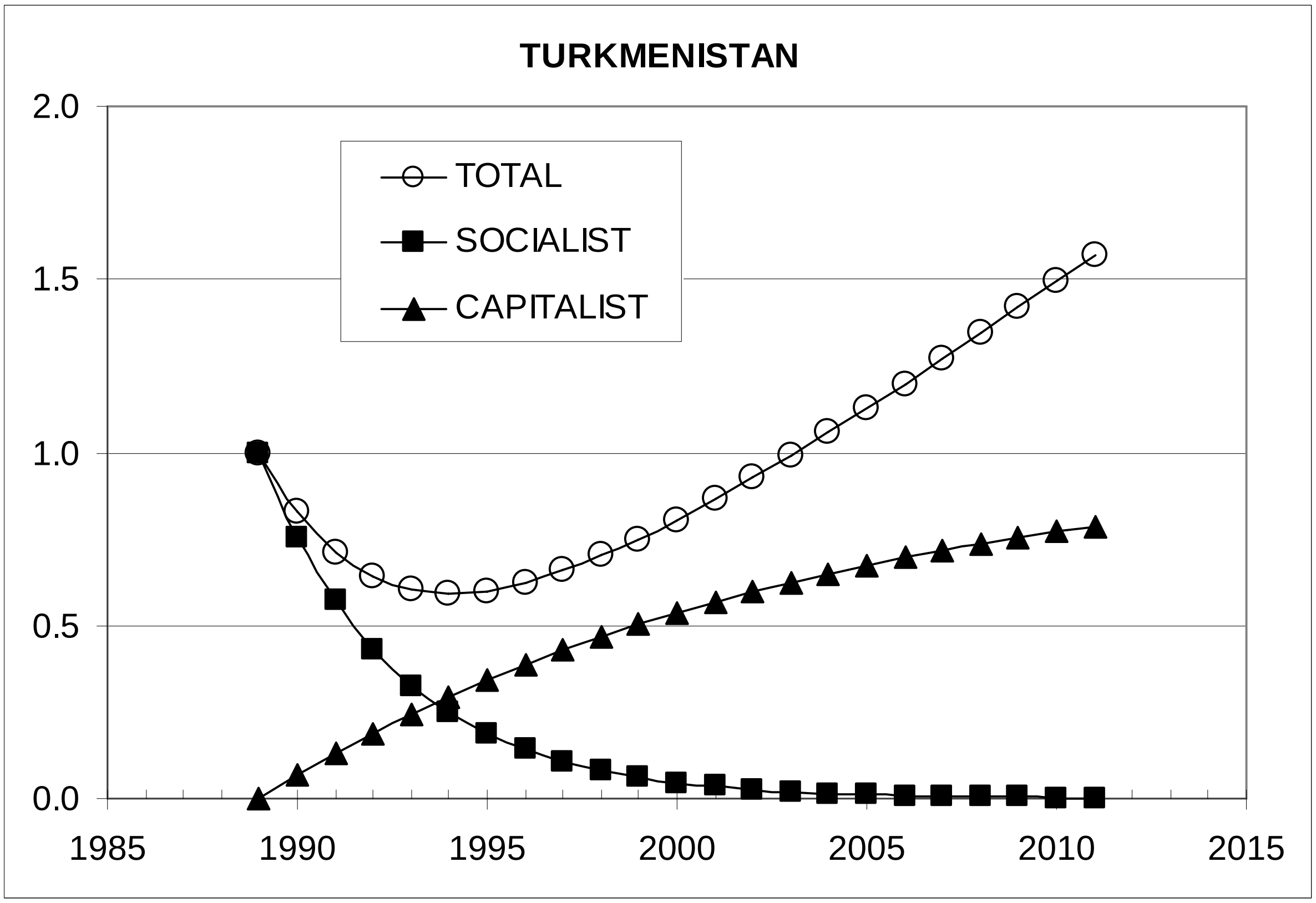

Fig. A17. Comparison of the observed and predicted transition process for the replacement of the socialist system with the capitalist system in Turkmenistan. Parameters of the model are indicated.

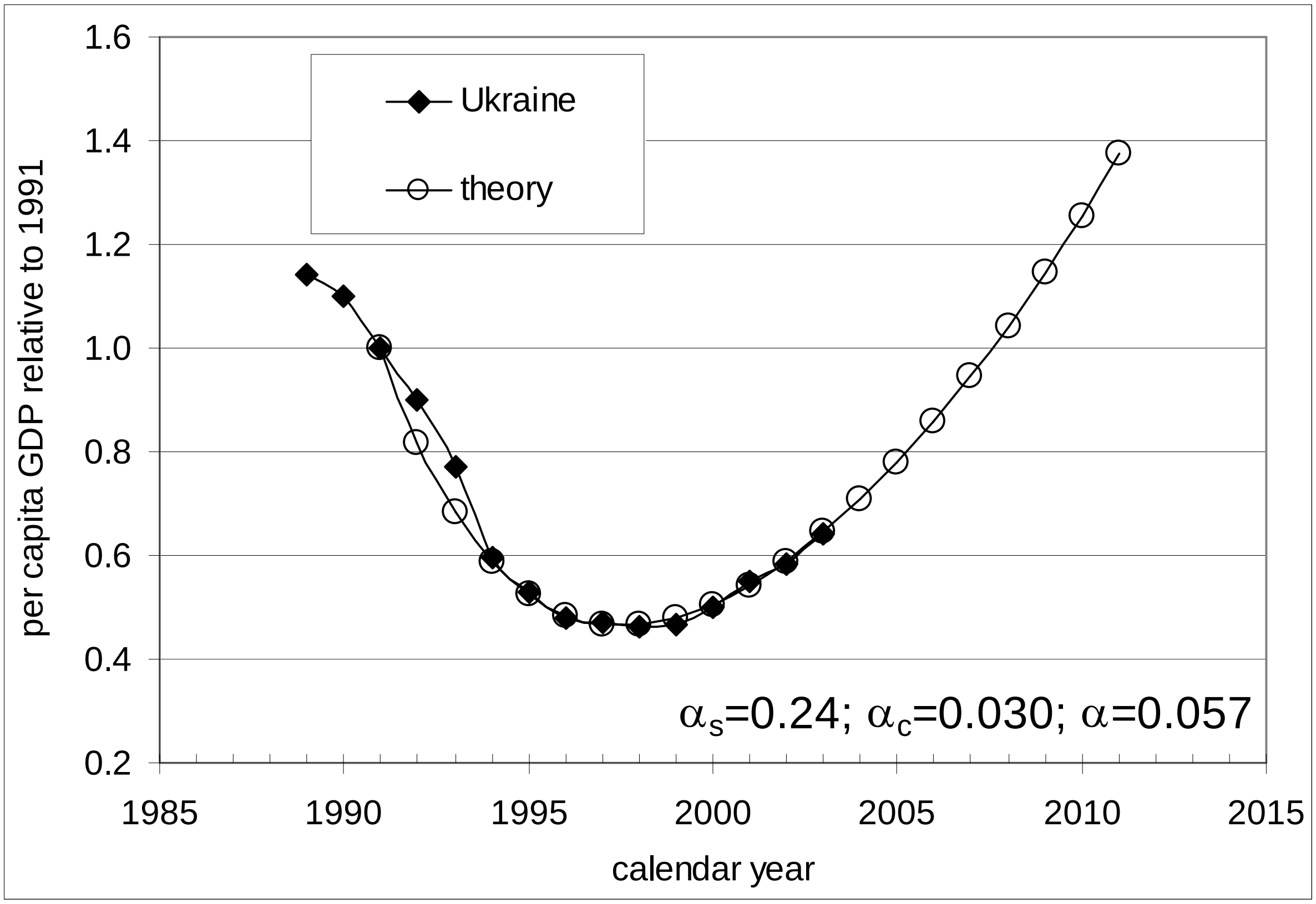

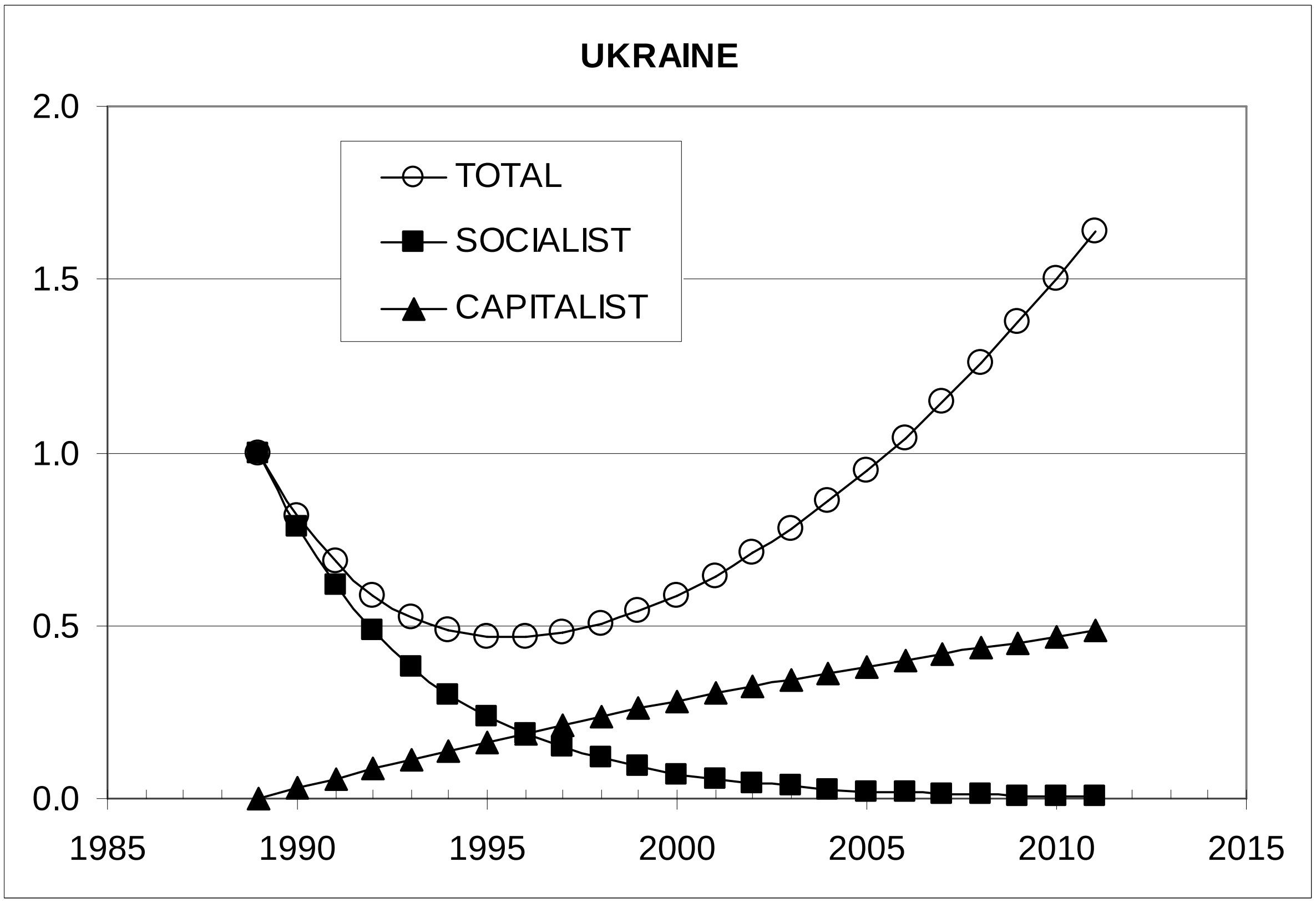

Fig. A18. Comparison of the observed and predicted transition process for the replacement of the socialist system with the capitalist system in Ukraine. Parameters of the model are indicated.

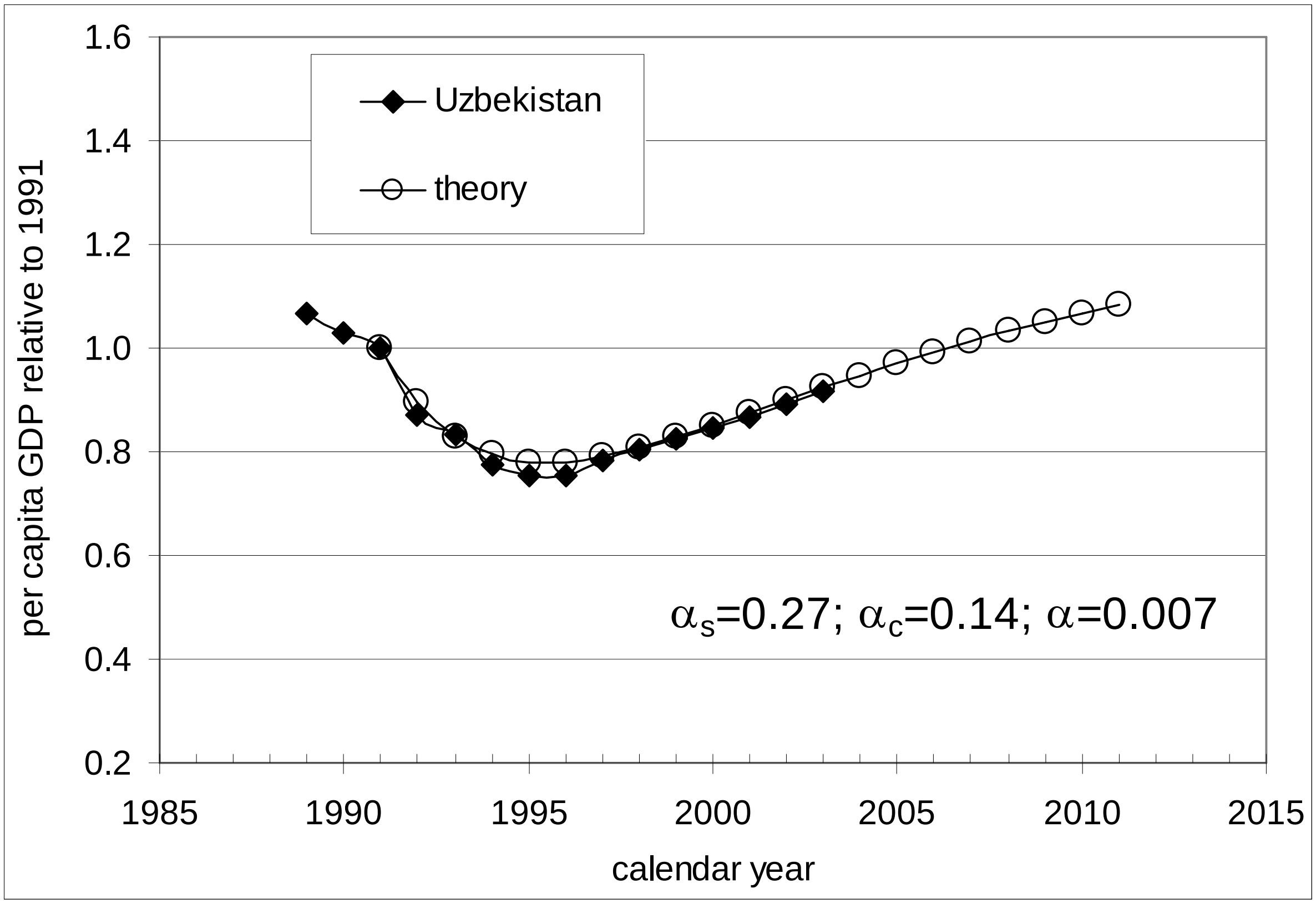

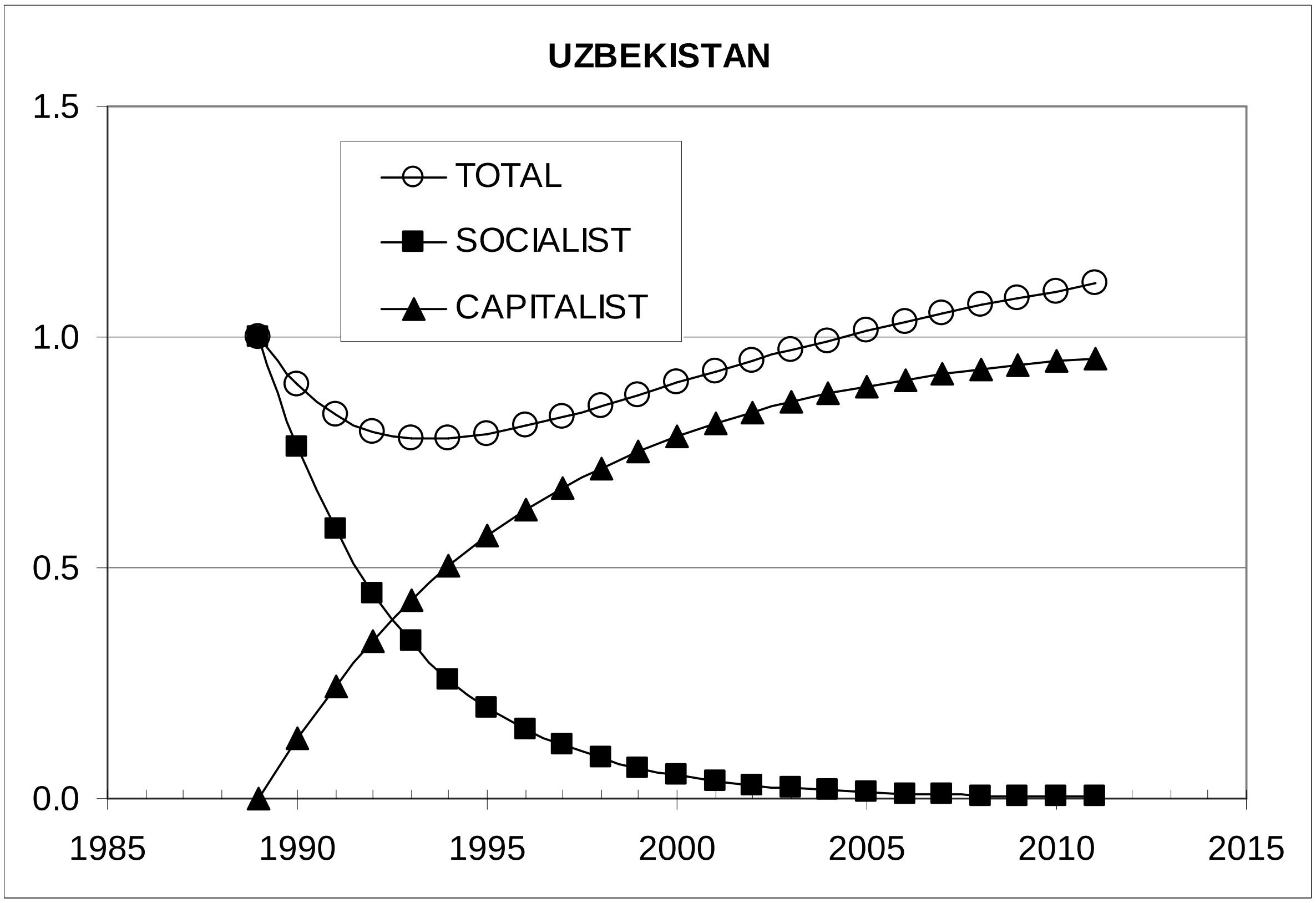

Fig. A19. Comparison of the observed and predicted transition process for the replacement of the socialist system with the capitalist system in Uzbekistan. Parameters of the model are indicated.

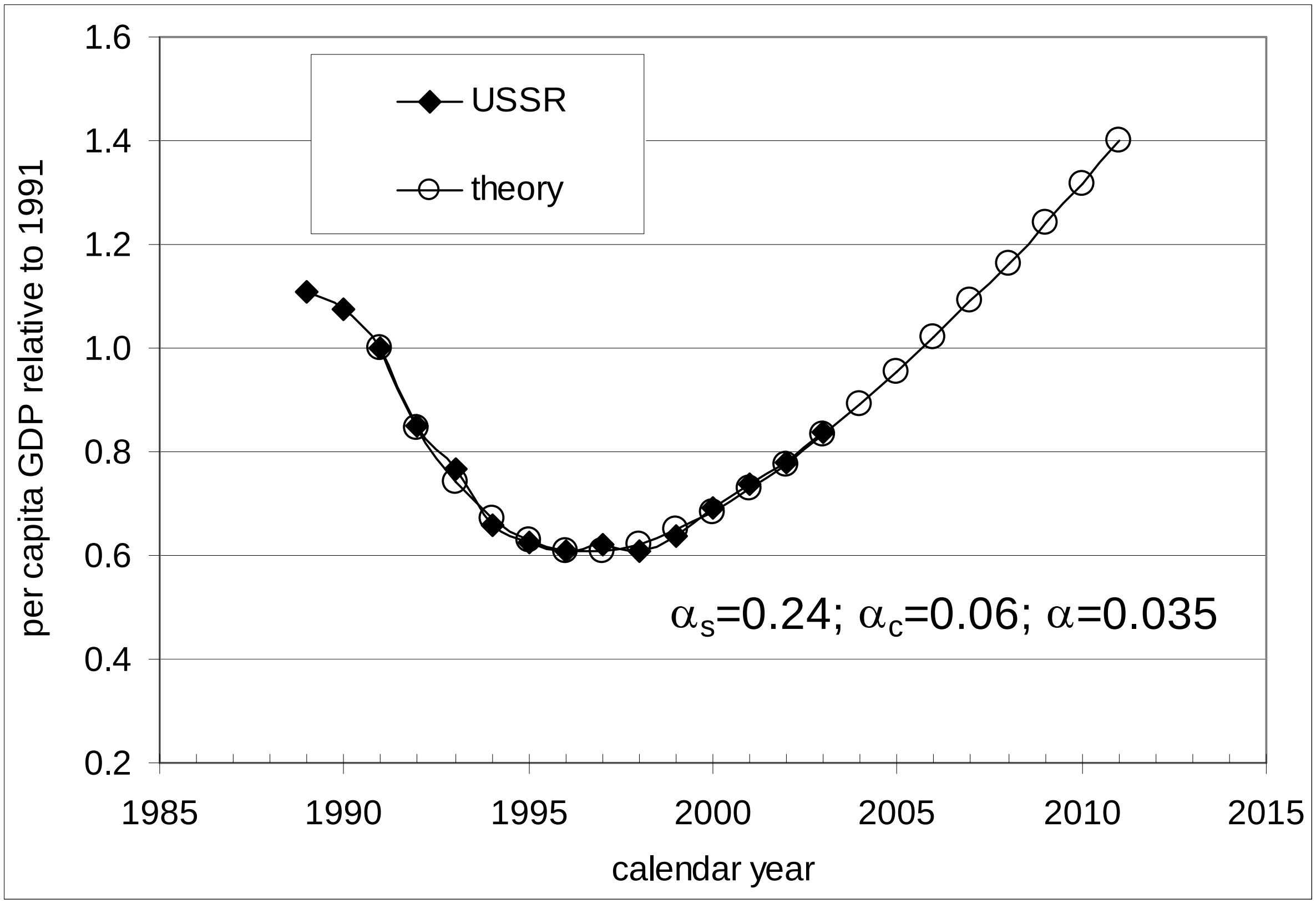

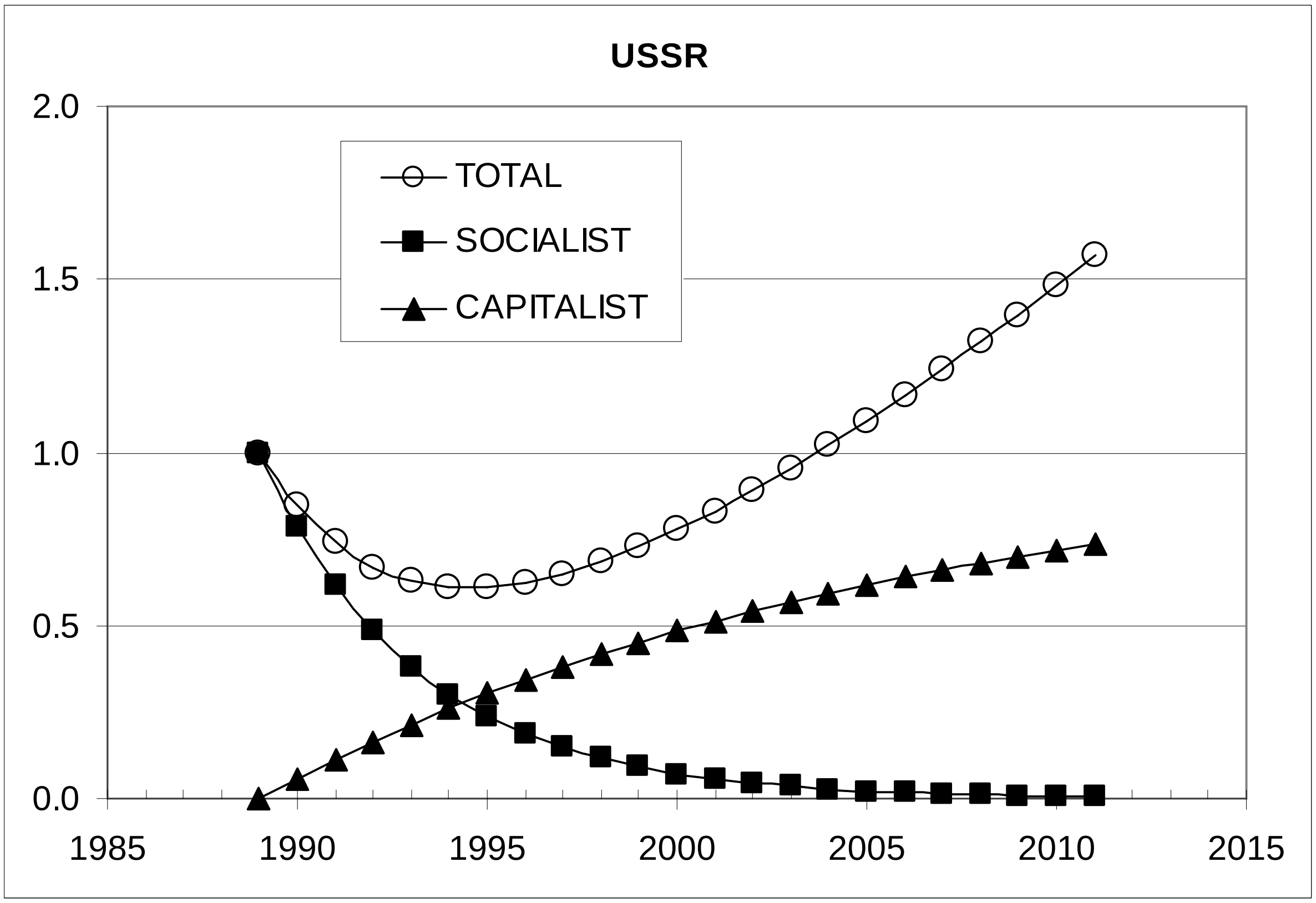

Fig. A20. Comparison of the observed and predicted transition process for the replacement of the socialist system with the capitalist system in USSR. Parameters of the model are indicated.